%% file: full_version.tex
\documentclass[conference]{IEEEtran}

\usepackage{mathtools}
\usepackage{alltt}
\usepackage{url}
\usepackage{bbm}
\usepackage{paralist}
\usepackage{listings}
\usepackage{color}
\usepackage{subfig}

\newcommand{\sys}[1]{\texttt{\textbf{Ver}}}
\newcommand{\topk}[1]{\textsc{FastTopK}}

\usepackage{listings}
\usepackage{multirow}
\usepackage{graphicx}
\usepackage[dvipsnames]{xcolor, colortbl}
\usepackage{soul}
\usepackage{marginnote}
\usepackage{balance}
\usepackage{enumitem}
\usepackage[linesnumbered, vlined, boxed, ruled]{algorithm2e}
\usepackage{multirow}
\usepackage{tabularx,ragged2e}
\usepackage{tcolorbox}

\newcommand{\ie}{i.e.,\ }
\newcommand{\eg}{e.g.,\ }

\newcommand{\bluecolor}[1]{{\color{Blue} #1}}
\newcommand{\tinyskip}{\vspace{3pt}}
\newcommand{\setting}{pathless table collection}

\newcommand{\D}{\mathcal{D}}
\newcommand{\mypar}[1]{\tinyskip\noindent\textbf{#1.}\xspace}
\newcommand{\myparwod}[1]{\tinyskip\noindent\textbf{#1}\xspace}

\newcommand{\F}{\mbox{Fig.\hspace{0.25em}}}

\newenvironment{myitemize}{%
\begin{itemize}[leftmargin=1em, itemsep=.1em, parsep=.1em, topsep=.1em,
    partopsep=.1em]}
{\end{itemize}}

\newenvironment{myenumerate}{%
\begin{enumerate}[leftmargin=1em, itemsep=.1em, parsep=.1em, topsep=.1em,
    partopsep=.1em]}
{\end{enumerate}}

\newenvironment{structure*}{\color{blue}\begin{myenumerate}}{\end{myenumerate}}

\sethlcolor{yellow}

\newtheorem{problem}{Problem}
\newtheorem{definition}{Definition}

\newcommand{\specialcell}[2][c]{%
  \begin{tabular}[#1]{@{}c@{}}#2\end{tabular}}

\newcolumntype{C}{>{\Centering\arraybackslash}X} 

\SetCommentSty{mycommfont}

\usepackage{tikz}
\usepackage{collcell}

\usepackage{etoolbox}
\usepackage{xstring}
\newtoggle{inTableHeader}
\toggletrue{inTableHeader}
\newcommand*{\StartTableHeader}{\global\toggletrue{inTableHeader}}%
\let\OldTabular\tabular%
\let\OldEndTabular\endtabular%
\renewenvironment{tabular}{\StartTableHeader\OldTabular}{\OldEndTabular\StartTableHeader}%

\newcommand*{\MinNumber}{0.0}%
\newcommand*{\MidNumber}{83} %
\newcommand*{\MaxNumber}{100.0}%
\usepackage{tikz}
\def\checkmark{\tikz\fill[scale=0.4](0,.35) -- (.25,0) -- (1,.7) -- (.25,.15) -- cycle;} 

\newcommand{\ApplyGradientnew}[4]{%
    \IfDecimal{#1}{%
      \edef\mynum{#1}%
      \ifdim #1 pt > #4 pt\relax
        \edef\mynum{#4}%
      \else
        \ifdim #1 pt < #2 pt\relax
          \edef\mynum{#2}%
        \fi
      \fi
      \ifdim \mynum pt > #2 pt
        \pgfmathsetmacro{\PercentColor}{max(min(100.0*(\mynum - #2)/(#4-#2),100.0),0.00)}%
        \xdef\PercentColorr{\PercentColor}
        \cellcolor{red!\PercentColorr}#1
      \else
        \pgfmathsetmacro{\PercentColor}{max(min(100.0*( \mynum -#2)/(#4-#2),100.0),0.00)}%
        \xdef\PercentColorr{\PercentColor}
        \cellcolor{red!\PercentColorr}#1%
      \fi	
    }{#1}
}

\newcommand{\ApplyGradient}[1]{%
    \IfDecimal{#1}{%
      \edef\mynum{#1}%
      \ifdim #1 pt > \MaxNumber pt\relax
        \edef\mynum{\MaxNumber}%
      \else
        \ifdim #1 pt < \MinNumber pt\relax
          \edef\mynum{\MinNumber}%
        \fi
      \fi
      \ifdim \mynum pt > \MidNumber pt
        \pgfmathsetmacro{\PercentColor}{max(min(100.0*(\mynum - \MidNumber)/(\MaxNumber-\MidNumber),100.0),0.00)}%
        \xdef\PercentColorr{\PercentColor}
        \cellcolor{red!\PercentColorr!yellow}#1
      \else
        \pgfmathsetmacro{\PercentColor}{max(100-min(100.0*(\MidNumber - \mynum)/(\MidNumber-\MinNumber),100.0),0.00)}%
        \xdef\PercentColorr{\PercentColor}
        \cellcolor{red!\PercentColorr}#1%
      \fi	
    }{#1}
}
 
\newcolumntype{R}{>{\collectcell\ApplyGradient}c<{\endcollectcell}}

\definecolor{Gray}{gray}{0.9}
\definecolor{maroon}{cmyk}{0,0.87,0.68,0.32}
\definecolor{red}{cmyk}{0, 0.68, 0.77, 0.09}
\definecolor{lightred}{cmyk}{0, 0.65, 0.76, 0.08}
\definecolor{orange}{cmyk}{0, 0.23, 0.74, 0.04}
\definecolor{LightCyan}{rgb}{0.88,1,1}
\definecolor{antiquewhite}{rgb}{0.98, 0.92, 0.84}

\newcolumntype{g}{>{\columncolor{Gray}}c}
\newcolumntype{m}{>{\columncolor{maroon}}c}

\begin{document}
%
\title{Ver: View Discovery in the Wild}

\author{\IEEEauthorblockN{Yue Gong\IEEEauthorrefmark{1},
 Zhiru Zhu\IEEEauthorrefmark{1}, Sainyam Galhotra, 
 Raul Castro Fernandez}
 \IEEEauthorblockA{Department of Computer Science,
 The University of Chicago\\
 Email: \{yuegong, zhiru, sainyam, raulcf\}@uchicago.edu}
 \IEEEauthorrefmark{1}Yue Gong and Zhiru Zhu contributed equally
 }

\maketitle

\input{sections/abstract}    


%

\input{sections/introduction}
\input{sections/background}

\input{sections/architecture_full}

\input{sections/vpresentation}
\input{sections/4c}

\input{sections/evaluation_full}

\input{sections/related_work}

\input{sections/conclusions}
\input{sections/acknowledgement}
\input{sections/appendix}
\bibliographystyle{IEEEtran}
\balance
\bibliography{main}

\end{document}

%% file: sections/abstract.tex
\begin{abstract}

We present \sys{}\footnote{code is available at https://github.com/TheDataStation/ver}, a data discovery system that identifies project-join views
over large repositories of tables that do not contain join path information, and even when input queries are inaccurate. \sys{} implements a reference architecture to solve both the technical (scale and search) and human (semantic ambiguity, navigating a large number of results) problems of view discovery. We demonstrate users find the view they want when using \sys{} with a user study and we demonstrate its performance with large-scale end-to-end experiments on real-world datasets containing tens of millions of join paths.



\end{abstract}

%% file: sections/introduction.tex
\section{Introduction}

The existence of large repositories of data such as those that originate from the combination of disparate databases~\cite{dataintegration}, data lakes~\cite{datalake}, open data portals~\cite{opendata}, and cloud repositories~\cite{lakehouse} has the upside of offering opportunities to find valuable data for tasks such as machine learning, reporting, and data analytics. The downside is the resulting data discovery problem: identifying a combination of datasets useful for the downstream task even when these reside in different databases. For example, a machine learning engineer may need a training dataset that requires combining a table in a database with the one sitting on the enterprise data lake. Large volumes of often incomplete and noisy data without any join information, which we call \emph{pathless table collections}, make solving data discovery difficult and thus hampers productivity.

There are several approaches to identify project-join views (PJ-views) over pathless table collections. Discovery systems such as Aurum~\cite{aurum}, Goods~\cite{halevy2016goods}, Auctus~\cite{auctus}, Juneau~\cite{zhang2019juneau}, Josie~\cite{zhu2019josie}, Table-Union~\cite{tableUnionSearch}, D3L~\cite{d3l} and libraries such as LSHEnsemble~\cite{zhu2016lsh}, and Lazo~\cite{fernandez2019lazo} help with identifying datasets that satisfy some relevance criteria whether via keywords~\cite{halevy2016goods} or programs~\cite{aurum}. Analysts can then combine the datasets to verify that it satisfies the view they need. Another approach is query-by-example (QBE) interfaces~\cite{zloof1975query, qbo} that lets users provide examples of the view they need. These can be adapted, with effort, to run on top of data discovery systems. Whether via keywords, programs, or QBE interfaces, the result of discovery queries over pathless table collections leads to ambiguous results that include semantically distinct results, duplicates or near-duplicates, and different versions of the data. This ambiguity further complicates identifying the right view. More generally, solving view discovery in the wild requires addressing the following challenges:





\begin{myitemize}

\item \textbf{Challenge 1. Noisy Queries.} Users provide queries that represent their best knowledge of the data in the form of keywords, programs, and examples. User-provided input may or may not appear in the table collection and they may be noisy and incorrect.

\item \textbf{Challenge 2. Noisy Join Paths.} Pathless table collections do not include join paths. Identifying true join paths automatically is impossible. We resort to identifying inclusion dependencies, a proxy to join paths, i.e., a join path implies an inclusion dependency but not the other way around. 

\item \textbf{Challenge 3. Large Number of Join Paths.} Large volumes of data result in a large number of join paths that require efficient data structures to represent and navigate them.



\item \textbf{Challenge 4. Noisy Result Views.} A large number of join paths implies many views may satisfy a user query. Such result views will be noisy due to ambiguity in the user query and noisy join paths. Concretely, there will be duplicate result views, views that are contained within each other, others that are complementary, and others that will show contradictory values for the same key.



\item \textbf{Challenge 5. Result View Navigation.} Result views will contain semantically ambiguous results, e.g., views with ``work address'' and ``home address''. Only users know the right context so the challenge is to elicit that context and use it to choose the view they need among all result views.



\end{myitemize}

In this paper, we introduce a \emph{reference architecture to identify PJ-views in the wild}. Discovering a PJ-view over pathless table collections requires understanding human preferences and requirements (which we refer to as a \emph{human} problem) and solving a \emph{technical} problem. The architecture divides the larger problem into smaller ones, each of which we tackle with a different component. Reference architectures help conceptualize problems and have been influential in advancing the field. For example, \cite{rdbms, inmemoryrdbms} for relational databases and \cite{openii, orchestra, datacivilizer, biggorilla} for data integration. The reference architecture we propose tackles the five challenges above. While challenges 1-3 are addressed by existing work, we introduce new techniques to address Challenge 4 and 5 in this paper, and demonstrate them as part of an implementation of the reference architecture. We make the following contributions:


\begin{myitemize}

\item An end-to-end system, \sys{}, that identifies PJ-views among tens of millions of join paths. \sys{} implements the reference architecture for QBE-based interfaces. We choose QBE because it permits users to declare the table they would like to find, even when they do not know examples and can only provide attribute names (Section~\ref{sec:architecture_full}). \sys{} relies on existing work~\cite{aurum} to address Challenges 1-3.

\item A \textbf{view presentation} component that helps humans to identify a good view among many. The approach uses a novel bandit-based approach to \emph{learn} user-specific discovery preferences, and it uses abduction-based reasoning to quickly narrow down the search space, reducing the number of human interactions (Section~\ref{sec:vpresentation}).

\item A \textbf{view distillation} component that automatically filters out views by classifying them into \textbf{4C categories}: \emph{compatible}, \emph{contained}, \emph{complementary}, and \emph{contradictory}. Besides helping with filtering, these categories produce information used by the view presentation component (Section~\ref{sec:vdistillation}).

\end{myitemize}

We conduct an IRB-approved user study (Institutional Review Board)~\cite{irb} to validate \sys{}'s approach to the human problem. We conduct thorough experiments on queries from open data repositories that lack join paths.


%% file: sections/background.tex


\section{Definitions and Problem Statement}
\label{subsec:problemdefinition}\label{sec:background}

$\mathcal{R}(A_1,\ldots,A_m)$ is a relation schema over $m$ attributes, where $A_i$ denotes the $i^{th}$ attribute.
A  table $D$ comprises a schema $\mathcal{R}(A_1,\ldots,A_m)$ and a set of tuples $T$ where each tuple $t\in T$ is a specific instance of the schema.

In practice, tables do not look like the ideal defined above because they may \emph{lack header information}, have \emph{ambiguous names} and contain \emph{dirty and noisy data}. More formally:



\begin{definition}[Noisy structured data]
A noisy data $D$ is characterized by an incomplete schema information $\mathcal{R}(A_1,\ldots,A_m)$ where $A_i=\phi$ for missing header values and tuples $T$ such that each tuple $t\in T$ contains at most $m$ values.
\end{definition}

In addition, a pathless collection may contain tables with contradictory values, \eg two census tables with different population counts for the same states of the country. Formally, two tables $D_i$ and $D_j$ contradict if the tables contain different values for the same key. We discuss the detection of contradictions in Section~\ref{sec:vdistillation}.


\begin{definition}[Pathless table collection] 
A pathless table collection contains a set of noisy tables $\mathcal{D} = \{D_1,\ldots, D_n\}$ where each $D_i$ is a noisy table and tables $D_i, D_j$ may contain contradictory values.
\end{definition}


\smallskip
\noindent \textbf{PJ-example-query.} A PJ-query (hereafter called query) contains (possibly noisy) example tuples of the desired output. The examples are a proxy to user's discovery requirements. The quality of examples depends on user's knowledge. Given a query $q$, there may be many tables that contain the input examples, and many combinations of these may satisfy $q$, resulting in a large number of candidate PJ-views.



\begin{definition}[Noisy query]
An input query $q$ is a noisy table consisting of $l$ example  tuples,  $\chi=\{\chi_1,\ldots, \chi_l\}$ where each $\chi_i$ is a noisy tuple denoting example values that are expected to be present in the desired output. The different columns in the examples $\chi$ are denoted by $\chi.A_i, \forall i\in\{1,\ldots, \tau \}$, where $\tau$ is the number of attributes in the input query. 
\end{definition}

PJ-views are constructed by joining datasets through keys. We first define a join path and then use it to discuss the effects of noise, followed by a formal problem statement.

\begin{definition}[Join path]
A join path $P$ is defined as an ordered set of noisy tables $P\equiv \{D_1,\ldots, D_t\}$ such that tables $D_j$ and $D_{j+1}$ join for all $j<t$ via a key column $k\in D_j,D_{j+1}$ forming a chain of join operations. 
\end{definition}

Joinable datasets can be identified in the presence of key information, which is generally absent in pathless scenarios. A PJ-view $V$ is the dataset formed after materializing the path $P$ followed by projection, i.e., choosing the relevant columns.



\begin{problem}[Project-Join view discovery over pathless table collections\label{prob:exploration}] Consider a \setting\ $\mathcal{D}$ and a query $q$ with examples $\chi$. The goal is to construct a minimal candidate set of PJ-views $\mathcal{D}'$ that satisfy the user requirements.
\end{problem}

Users may require the PJ-view to contain all examples $\chi$, or any of them, depending on user and application. Formalizing these requirements is outside the scope of this work.

%% file: sections/architecture_full.tex
\section{A Reference Architecture}
\label{sec:architecture_full}


\myparwod{Why a Reference Architecture?} From a systems engineering perspective, a reference architecture is the materialization of a ``divide and conquer'' strategy that splits complex engineering problems into smaller parts. Thus, a reference architecture describes a collection of components and their interactions. In making a reference architecture concrete, we state our understanding of the problem and represent it as a concrete artifact that the community can scrutinize and improve.

\subsection*{Design Overview}

Finding a PJ-view in the wild requires solving a \emph{human} and a \emph{technical} problem. We present a reference architecture for view discovery that contains components targeting both \emph{human} and \emph{technical} problems. Along with each component, we discuss implementation options, including those made by \sys{}. We also briefly discuss two novel components, \textsc{View Distillation} and \textsc{View Presentation}.



\begin{figure}[t]
  \centering
  \includegraphics[width=0.9\linewidth]{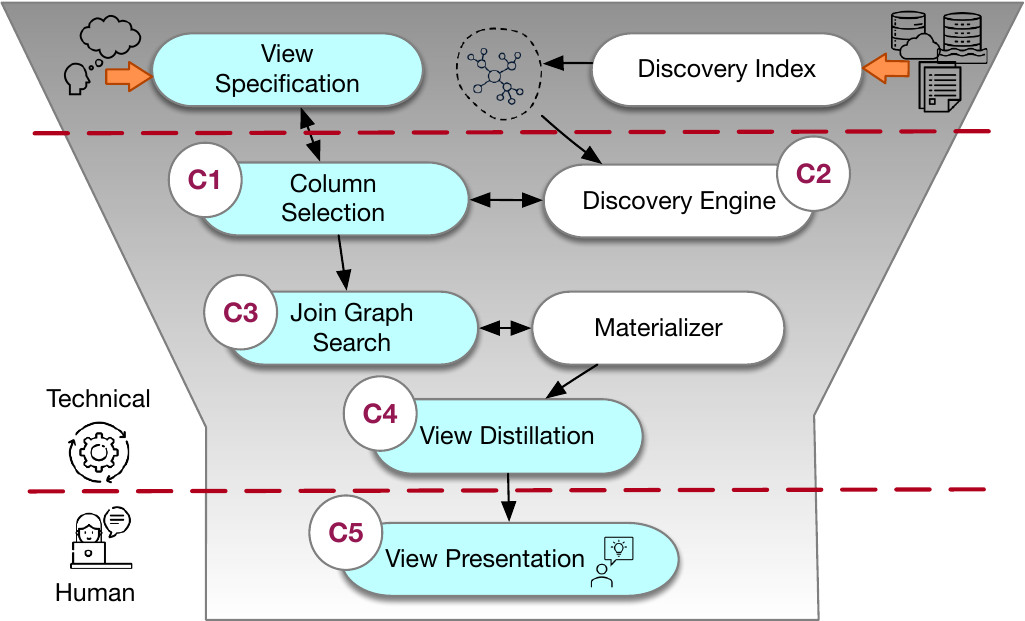}
  \caption{\sys{} Reference Architecture. The funnel illustrates the progressive reduction of data as it flows downstream.}
  \label{fig:architecture}
\end{figure}

\mypar{Overview} Algorithm~\ref{algo_dod} shows pseudocode (human components are {\color{Blue}highlighted in blue}) and \F\ref{fig:architecture} shows \sys{}'s architecture inside a funnel, denoting the gradual reduction of views as data flows downstream. \sys{} builds a discovery index offline. 


      

\mypar{\textsc{Discovery Engine and Index Creation (technical)}} \emph{This component builds indices over pathless table collections: i) a join path index, which can be approximate; ii) retrieval indices over table names, values, attribute names and column similarity. The indices are available online, via the Engine's API to other components. (\textbf{Challenge 2}).}



The indices can be built using state-of-the-art methods such as Aurum~\cite{aurum} (that \sys{}'s implementation uses), Auctus~\cite{auctus}, JOSIE~\cite{zhu2019josie}, LSHEnsemble~\cite{zhu2016lsh}. After the indices are built, users design and submit queries via a \textsc{view-specification} component (line 1).

      

\mypar{\textsc{View Specification (human)}}
Discovery interfaces include spreadsheet-style, keyword search, APIs, natural language, and combinations of these. The reference architecture supports these interfaces via the \textsc{View-Specification} component. For QBE-based interfaces, as implemented by \sys{}, the input is a set of examples, $\chi$, and the output of this stage is a set of example attributes and values. \textsc{View-Specification} may interact with users when it detects the provided examples are ambiguous, or to offer examples to choose from.


Next, the \textsc{column-selection} component selects the subset of tables containing user-provided examples, $\chi$ (lines~\ref{dodline3}-\ref{dodline7}).


      

\mypar{\textsc{Column Selection (technical)}} \emph{This component must be designed to identify relevant data even when the input query is noisy, addressing \textbf{Challenge 1} above. The output of this component is a collection of candidate tables and columns.}


\begin{algorithm}[t]
    \scriptsize
    \SetKwFunction{Union}{Union} \SetKwInOut{Input}{Input}\SetKwInOut{Output}{Output}
	
	\Input{Pathless table collection $\D$, Discovery Index $I$} 
	\Output{PJ-views $\mathcal{V}$}
	 \BlankLine 
	 \bluecolor{$\chi \leftarrow\textsc{View-Specification}(\mathcal{D})$}\\
	 $\textsc{Cand}\leftarrow \phi$\\
	 \For{$\chi.A_i\in \textsc{Columns}(\chi)$\label{dodline3}}{ 
	    $\textsc{Cand}(A_i) \leftarrow \textsc{Column-Selection} (\chi.A_i, \D, I)$\\
	    \If{\textsc{Mode}==Interactive}{
	   \bluecolor{ $\textsc{Cand}(A_i)\leftarrow $ Query $\textsc{Cand}(A_i)$} \\
	    }
	    $\textsc{Cand}\leftarrow	    \textsc{Cand}\cup \textsc{Cand}(A_i)$\label{dodline7}
	}
	$\mathcal{V}_{PJ} \leftarrow \textsc{Join-Graph-Search} (\textsc{Cand},\chi)$\label{dodline8}\\
	$\mathcal{S} \leftarrow \textsc{View-Distillation} (\mathcal{V}_{PJ})$\label{dodline9}\\
	\If{\textsc{Mode}==Interactive\label{dodline10}}{
	\bluecolor{$\mathcal{V} \leftarrow \textsc{View-Presentation}(\mathcal{V}_{PJ},\mathcal{S})$} 
	}
	\Else{
	$\mathcal{V} \leftarrow $ Rank $\mathcal{V}_{PJ}$ based on overlap score\label{dodline13}
	}
	 \Return $\mathcal{V}$
 	 	  \caption{\sys{} Design Overview 
 	 	  }
 	 	  \label{algo_dod} 
\end{algorithm}

The candidate columns are processed by the \textsc{join-graph-search} component (line~\ref{dodline8}) to enumerate and materialize candidate PJ-views (addresses Challenge 3). The candidate view search space reduces, as shown in \F\ref{fig:architecture}.


      

\mypar{\textsc{Join Graph Search (technical)}} \emph{Given a set of candidate tables, an input query, and the discovery index providing join paths, this component identifies all \emph{join graphs} that, when materialized, produce candidate PJ-views. The main goal of this component is to address the large join path space (\textbf{Challenge 3}). To materialize candidate PJ-views, the \textsc{join-graph-search} component uses a \textsc{Materializer}, a data processing component with the capacity to execute PJ queries.}


\textsc{Join-Graph-Search} returns many \emph{candidate views}. Ranking the views is hard because of users' differing search criteria. The \textsc{view-distillation} component (line~\ref{dodline9}) summarizes the candidate views, further reducing the view search space as shown in \F\ref{fig:architecture}.

      

\mypar{\textsc{View Distillation (technical)}} \emph{This component computes \emph{categories} from the candidate PJ-views that include redundancy and containment in the views, as well as opportunities for unioning views and more. Some categories can be used to distill/summarize the views (\textbf{Challenge 4}). Others are shared with the downstream component.}

\textsc{view-presentation} 
receives the distilled views. It can \emph{rank} the views and return top-1 (for a full automated mode) or it can leverage the categories computed by \textsc{view-distillation} to help users find the right data (lines~\ref{dodline10}-\ref{dodline13}).


      

\mypar{\textsc{View Presentation (human)}} \emph{\textsc{View-presentation} uses different question interfaces to elicit information from users via data questions. The questions are designed to narrow down the space until users find the desired view (\textbf{Challenge 5}). The component chooses what questions to ask, sequentially, using a bandit-based approach.}



Next, we present \textsc{View-Presentation} (Section \ref{sec:vpresentation}), \textsc{View-Distillation} (Section \ref{sec:vdistillation}). We also introduce the implementation details of \textsc{Discovery-Index}, \textsc{Column-Selection} and \textsc{Join-Graph-Search} in Appendix.

%% file: sections/vpresentation.tex
\section{View Presentation}
\label{sec:vpresentation}

Ideally, a query results in one PJ-view. In practice, ambiguity, redundancy, erroneous join paths, and large table repositories mean there may be hundreds of result views. \sys{} uses novel techniques to reduce the number of views automatically (Section~\ref{sec:vdistillation}), but there is a limit to automation. Semantic ambiguity requires involving users to obtain the final view. The \textsc{view-presentation} component analyzes the views and generates questions that, when answered, help rank and select views. For example, this component will ask a user if they want ``home address'' or ``work address'' in their output when it detects both in the views. By asking questions, users learn more about the schemas and datasets available, and thus refine their preferences and discovery needs. A key challenge is that different users may be able to respond to different questions and that their preferences evolve as they interact with the system. \sys{}'s \textsc{view-presentation} component is based on two design principles that cater to varied user needs.



\noindent$\bullet$~\textbf{Mixed-Initiative Interface Design:} Lack of knowledge about available datasets inhibits users from effectively querying the identified set of views. However, users can answer questions about their desired view. Each user has a different understanding of the requirements and finds different interfaces to be more appropriate depending on that. Therefore, users may need different interface designs to answer questions. For example, some users could recognize phone number from the area code, while others might look for a pattern across values. Because different users will be able to answer different questions, \sys{} supports different question interfaces, each asking a different question type. This is motivated by previous mixed-initiative designs~\cite{vartak2014seedb,horvitz1999principles}. Besides, \sys{} learns which interface to offer to a user according to their previous actions. 

\noindent$\bullet$~\textbf{Adapt to evolving users' knowledge:} Users' understanding of their discovery need evolves as they interact with questions and learn about the schemas and data contained in the repository. \sys{} is designed so users can change their mind about previously answered questions, thus using newly acquired knowledge, without forcing users to start afresh.


\mypar{Question Interface} To cater to the diverse preferences of users, \textbf{\texttt{Ver}} considers the following collection of different interface designs.


\noindent$\bullet$~\textbf{Dataset} \textbf{interface}: This interface shows users a candidate view to check if it satisfies user's requirements.

\noindent$\bullet$~\textbf{Attribute} \textbf{interface}: This interface shows users an attribute and asks if it should be present in the desired output.

\noindent$\bullet$~\textbf{Dataset} \textbf{Pair}: This interface shows users a pair of views and asks them to pick one. This interface is specifically designed to leverage 4C categorization of views (Section~\ref{sec:vdistillation}).

\noindent$\bullet$~\textbf{Summary} \textbf{interface}: This interface summarizes a collection of views and checks if it is relevant for the desired output. We use a wordcloud to visualize the summary. 
     

At every iteration, there are two choices to make: i) what interface design to choose (i.e., how to show the user the question); ii) what prioritization strategy to use to choose the question to show on the chosen interface. For example, if the algorithm chooses the attribute interface, then it could show an attribute similar to the input query or one that is different from others previously shown. We implement two prioritization strategies to order questions: i) distance of the question from the input query; ii) distance of dataset schema to input query. \textbf{\texttt{Ver}} supports other interface designs and prioritization strategies. Finally, users can always skip any question and \textbf{\texttt{Ver}} adapts to their responses. We discuss the relative usefulness of these interfaces in Section~\ref{sec:userstudy}.


\textbf{\texttt{Ver}}'s view presentation addresses the following problem

\begin{problem}
Given a collection of views $\mathbbm{V}$ and access to a user who answers questions through interfaces $\mathbbm{I}$,
 prioritize questions to identify the desired view while minimizing the number of queries to the user.
\end{problem}

The \textsc{view-presentation} component is designed to help users interact with the collection of candidate views and effectively navigate the result views. We now present the key insights we use to solve the problem and \textbf{\texttt{Ver}}'s view presentation algorithm.

\subsection{Bandit-Based View Presentation Algorithm} A key design principle of the \textsc{View-Presentation} component is to not prune any views unless specifically discarded by the user. Instead, the \textsc{View-Presentation} component ranks the views, giving users the ability to revisit their choices as their knowledge evolves. It must balance the need for asking informative questions that help narrow down views with questions that the specific user may be able to answer. Merely learning first what question interface the user prefers (exploration stage) and then asking questions based on that interface alone (exploitation stage) would be brittle to users' changing knowledge and preferences. The algorithm is based on two insights: i) learning the best interface for a given user can be modeled as a bandit problem and; ii) the reward in the bandit problem should be based on a question's potential to reduce the number of candidate views.

\mypar{Bandit-Based approach} A multi-arm bandit algorithm naturally models probabilistic user preferences. Each question interface is an arm, a question-answer pair is pulling an arm, and the reduction in candidate views after the answer is the reward. We design the algorithm off Exp3~\cite{auer2002nonstochastic} because i) it does not make assumptions about the reward distribution; ii) the expected reward is represented as the arm's weights and; iii) has provable guarantees. Exp3 uses an exponential growth function to adapt weights of arms that obtain a positive reward, and all arms start with the same weight and are considered independent from each other. We improve on this behavior by leveraging knowledge of the user's expected reward.


\mypar{Question's reward} The reward of a question $q$ is its expected information gain, defined as the maximum number of irrelevant views that are pruned if the user answers $q$. Information gain becomes the reward of the bandit formulation and thus it guides the questions that \textsc{view-presentation} asks users.


\IncMargin{1em}
\begin{algorithm}\SetKwFunction{Union}{Union}
\scriptsize
\SetKwInOut{Input}{Input}\SetKwInOut{Output}{Output}
	\Input{Candidate set of views $\mathbbm{X}$, Set of question interfaces $\mathbbm{I}$, Exploration factor $\gamma$  } 
	\Output{Set of required views $\mathbbm{S}$.}
	 \BlankLine 
	 
	 $\mathbbm{S} \leftarrow \mathbbm{X}$ \\
	 \While{$j \in \{1,2,\ldots, T\}$}{
	 	 \For{ $I\in \mathbbm{I}$ }
	 {
	 \tcc*[h]{{\scriptsize Iterate over interfaces to estimate probability of selection}}\\
	    $r(I)\leftarrow$ Estimate likelihood to answer question with interface $I$\\
	    $\chi(I)\leftarrow$ Calculate info gain if the question using $I$ is answered\\
	    $w(I)\leftarrow r(I) \times \chi(I)$
	 }
	 \tcc*[h]{{\scriptsize Normalize to calculate probability}}\\
	 $p(I) \leftarrow (1-\gamma) \frac{w(I)}{\sum_{I \in \mathbbm{I} w(I)}} + \gamma / |\mathbbm{I}|, \forall I\in \mathbbm{I}$\\
	 $I_c$ $\leftarrow $ Draw randomly according to distribution $p$\\
	 response $\leftarrow $ Query the user using $I_c$\\
	 Update $r(I_c)$\\
	 \If{response $\neq$ \texttt{Skip}}{
    	 reward, $\mathbbm{S}$ $\leftarrow $ Use user response to remove irrelevant views from $\mathbbm{S}$ and update ranking\\
	 }
	 }
	 \Return $\mathbbm{S}$
 	 	  \caption{\textsc{View-Presentation}}
 	 	  \label{algo_view_presentation} 
 	 \end{algorithm}

Concretely, the probability of choosing an arm is as follows:


\begin{align}\nonumber
    p(I) = (1-\gamma) \frac{w(I)}{\sum_{I \in \mathbbm{I}} w(I)} + \frac{\gamma}{|\mathbbm{I}|}
\end{align}

where $\mathbbm{I}$ denotes the set of different question interfaces. $\gamma$ determines the probability of exploring a random arm, while ignoring the expected reward, and $w(I)$ denotes the estimated value of the expected information gain on question interface $I$. Choosing $\gamma=1$ is equivalent to an exploration strategy that chooses a random arm for every question while $\gamma=0$ chooses an arm that relies on expected reward estimation for each question. The expected information gain $w(I)$ 
of an interface $I$ is $r(I) \times \chi(I)$, where $r(I)$ is the probability that a user would answer the question with interface $I$, and $\chi(I)$ is the maximum reduction in candidate set size if the question using $I$ is answered. Initially, users have not answered any questions and the estimates of expected gain are not accurate. Therefore, the approach is bootstrapped with the exploration strategy until $O(\log |\mathbbm{I}|)$ questions have been asked for each interface. The estimated user behavior is then used to transform to a bandit-based approach. 

\mypar{Performance Guarantees} Theoretically, using Chernoff bound~\cite{chernoff1952measure} we can show that $O(\log |\mathbbm{I}|)$ questions per interface $I$ yield an accurate estimate of the $r(I)$ with a probability of $1-\frac{1}{|\mathbbm{I}|^2}$. Using prior results on the maximum coverage problem~\cite{nemhauser1978analysis}, greedily choosing the question that maximizes information gain is the best approximation of the optimal strategy. Therefore, the accurate reward estimation ensures effective interaction using a multi-arm bandit approach for \textsc{view-presentation}.
 
\subsection{The View Presentation Algorithm}

Algorithm~\ref{algo_view_presentation} first initializes the candidate set of views $\mathbbm{S}$ (line 1) and iteratively queries the user until $T$ iterations (lines 2--12). $T$ is used to denote any stopping criterion, which could be when the user ends the session or a pre-defined parameter. In each iteration, the multi-arm bandit based approach estimates the expected reward of each arm to calculate the probability distribution of choosing each arm (lines 3--7). The question interface is chosen according to this distribution. After choosing the interface $I_c$, a question that has the maximum information gain  $\chi(I)$ is asked to the user. User's response is subsequently used to either update $r(I)$ or the candidate set of views.
 
\mypar{Ranking Views} Given a collection of questions $Q$ to the user, an expected utility score is calculated for each view to rank them. Mathematically,  the utility score of a dataset $D$ is a weighted sum of the view's utility according to each question:

%
{%

\begin{align*}
 \sum_{Q_i\in Q} s_{Q_i} \times \left( P(D \text{ satisfies user needs} | Q_i =\checkmark)\cdot P(Q_i=\checkmark)\right)
\end{align*}

}%

 where $Q_i=\checkmark$ denotes that the question $Q_i$ is answered correctly and $s_{Q_i}$ is $1$ if $D$ is considered to satisfy user requirements by $Q_i$'s response, $-1$ if $D$ is considered irrelevant by $Q_i$'s response and $0$ otherwise. The probability P(D{ satisfies user needs} $|$ {$Q_i = \checkmark$}) is inversely proportional to the number of views that  $Q_i$ captures and the probability that a user would answer a question is used as a proxy estimate for P($Q_i=\checkmark$). Note that this score is calculated only for candidate views that are not pruned by user's responses.
 

%% file: sections/4c.tex
\section{View Distillation}
\label{sec:vdistillation}

\textsc{View-Distillation} consists of two parts: i) categorizing pairs of result views (i.e., candidate views); ii) applying a distillation strategy to reduce the number of result views.

\mypar{4C Categories} Result views with the same schema are classified into the following categories:

\begin{definition}[Compatible view pair]
Two candidate views, $V_1$ and $V_2$, are compatible (denoted by $V_1\equiv V_2$) if they have the same set of rows, $(V_1 \cap V_2) =  V_1=V_2$.
\end{definition}

\begin{definition}[Contained view pair]
A view, $V_1$, contains another view, $V_2$, when $V_2\subset V_1$, that is, when all rows of $V_2$ are contained in $V_1$.
\end{definition}

 A pair of views may be  \emph{Complementary} or \emph{Contradictory}, based on their candidate keys:


\begin{definition}[Candidate key]
A candidate key, $\mathcal{K}(V)$, is a set of attributes in an output view, $V$, that uniquely identify each row in $R\in V$. 
\end{definition}

Now we define complementary and contradictory pairs.

\begin{definition}[Complementary view pair]
Two views, $V_1$ and $V_2$ are complementary if the two views have the same candidate key, $K(V_1)=K(V_2)$, and the views have overlapping rows $V_1\cap V_2 >0$ but are neither contained nor compatible.
\end{definition}

\begin{definition}[Contradictory view pair] Two views, $V_1$ and $V_2$ are contradictory if the two views have the same candidate key, $K(V_1)=K(V_2)$, and a key value yields different rows in $V_1$ and $V_2$. 
\end{definition}
\mypar{Note} We categorize a pair of views as contradictory or complementary with respect to a candidate key. Therefore, views $V_1$ and $V_2$ may be contradictory with candidate key $k_1$ and complementary with key $k_2$. Our \textsc{view-distillation} component exposes all candidate relationships for further downstream processing.

The first step in \textsc{view-distillation} is to identify and label pairs of candidate views with one of the 4C categories. Then, a distillation strategy automatically prunes views based on their 4C category. Views are nodes in a graph. An edge is labeled with the category of the nodes it links. More formally:


\begin{problem}
Given a collection of views $\mathbbm{V}$, identify a labelled graph $G$ where the irrelevant views are pruned and edges that can be categorized as 4C are labelled accordingly.
\end{problem}

{
\IncMargin{1em}
\begin{algorithm}\SetKwFunction{Union}{Union}
\scriptsize
\SetKwInOut{Input}{Input}\SetKwInOut{Output}{Output}
\Input{$\mathbbm{V}$, collection of views}
\Output{ $G$, graph with edges categorized as 4C}
\BlankLine
 $G\leftarrow \textsc{Add-Nodes}(\mathbbm{V})$,$\mathbbm{V}\leftarrow \textsc{Identify-Keys}(\mathbbm{V})$\\
 $\mathbbm{V} \leftarrow \textsc{Schema-Based-Blocks}(\mathbbm{V})$\\
 \For{$\mathcal{V} \in \mathbbm{V}$}{
 \For{$1\leq i,j\leq |\mathcal{V}|, i>j$}{
    $V_i\leftarrow \mathcal{V}[i], V_j\leftarrow \mathcal{V}[j]$\\
    \tcc*[h]{\scriptsize Iterate over the views, compare rowwise hashes $H$}\\
    \If{$H[V_i] = H[V_j]$ }{
        $G[(V_i,V_j)]=$ Compatible\\
        $\mathcal{V}\leftarrow \mathcal{V}\setminus\{V_i\}$\\
        }
    \ElseIf{$H[V_i] \subset H[V_j]$ \text{ or } $H[V_j] \subset H[V_i]$   }{
    {   
        $G[(V_i,V_j)]=$ Contained\\
        $\mathcal{V}\leftarrow \mathcal{V}\setminus\{V_i,V_j\} \cup \{V_i\cup V_j\}$\\
    }}
    \ElseIf{$H[V_i] \cap H[V_j]\neq \phi$ }{
    {
    \tcc*[h]{\scriptsize Initialize any overlapping view pair as complementary}\\ 
        $G[(V_i,V_j)]=$ Complementary\\
    }
    }
    
 }
 \tcc*[h]{\scriptsize Second phase: Identify contradictions}\\
 $I\leftarrow \textsc{Index}(\mathcal{V})$\\
  \For{$k\in I.\texttt{keys}()$}{
    $\mathcal{C} \leftarrow \textsc{Group} (I[k])$\\
    \tcc*[h]{\scriptsize Identify pairs in different groups}\\
    \For{$(V_i,V_j) : V_i,V_j\in I[k],~ \mathcal{C}[V_i] \neq \mathcal{C}[V_j]$}{
    $G[(V_i,V_j)]=$ Contradictory
    }
  }
 }
 \caption{View Distillation}
 \label{alg:4c_main}
\end{algorithm}
}

\textbf{\texttt{Ver}}'s \textsc{view-distillation} implementation uses the following insights to classify views into 4C categories and construct the graph $G$. First, it uses the transitivity property to not compare any pair of views whose categorization can be inferred from prior comparisons. For compatibility, if $V_1\equiv V_2$ and $V_2\equiv V_3$, then $V_1\equiv V_3$. And for containment, \textbf{\texttt{Ver}} maintains the largest view for categorization, i.e. if $V_1\subseteq V_2$, then \textbf{\texttt{Ver}} distills out $V_1$ and keeps $V_2$ as $V_1$'s representative. Second, it partitions candidate views into \textsc{Schema-Based-Blocks}. This ensures that pairs of views are compared only if they share the same schema. Third, it hashes each view using a row-wise hash function (say $H$), i.e. $H(V)$ maps $V$ to a set of values where each value corresponds to a different row. This hash map helps to efficiently find compatible and contained pairs of views. Fourth, it identifies approximate key columns~\cite{key1,key2} and constructs an inverted index that maps each value in a key column to the corresponding rows and views that contain that value. This index helps to identify rows that have contradictory values and hence contradictory views.

Algorithm~\ref{alg:4c_main} presents the pseudocode of \textbf{\texttt{Ver}}'s \textsc{view-distillation}. First, it initializes a graph $G$ where each view is added as a node, identifies keys in each view (line 1) and partitions the collection of views $\mathbbm{V}$ into different blocks based on their schema (line 2). These schema-based blocks are processed sequentially to populate $G$ with 4C categories (line 3). The categorization process operates in two phases. The first phase (line 4-13) hashes all views (hash value of a view $V$ is denoted by $H(V)$) and compares hashed values to check containment and compatibility (line 6-11).  A pair of views $V_i$ and $V_j$ that overlap and have the same key but are not contained or compatible are marked as complementary (line 12-13). These pairs are later updated to be contradictory if the second phase identifies any contradictions.  All previously described comparisons are performed on the hash of each view. The hash function maps each view to a set of values, where each value in the set corresponds to a row in the view and we employ a cache to not hash any view multiple times. The second phase constructs an inverted index over the values in the key column(s) of each view in $\mathcal{V}$. This index maps each key value (say $k$) to a list of rows that contain the value $k$ (denoted by $I[k]$). \texttt{\textbf{Ver}} iterates over all values of the key column and identifies contradictions among the rows that contain the value (line 16-18). Specifically, it groups all duplicate rows that contain a key  value $k$ together (line 16) and pairs of views not in the same group are labeled contradictory (line 17-18).

\mypar{Distillation Strategy} Algorithm~\ref{alg:4c_main} merges distillation with graph construction. It applies a distillation strategy that deduplicates compatible views and keeps the largest contained view. Alternative strategies can be implemented based on the target use. This strategy helps reduce the search space of views that \textsc{View-Presentation} component needs to consider.


\mypar{Complexity Analysis} A crucial step of Algorithm~\ref{alg:4c_main} is hashing, which requires $O(n)$ time, where $n$ is the total number of candidate views. Other than that, \textbf{\texttt{Ver}} partitions the set of views into different schema blocks and compares views within a block. In the worst case, it may compare hashes of all pairs of non-compatible views within a block to check containment, i.e. requiring total complexity of $O(n + \alpha \Gamma^2)$, where $\alpha$ denotes the number of distinct schemas and $\Gamma$ denotes the maximum number of distinct views sharing the same schema. However, the distillation property of keeping the largest contained view helps reduce complexity in practical scenarios (median reduction ratio of more than $18\%$).  For contradiction and complementary categorization, calculation of key is the most time-consuming step, which requires processing all views. The subsequent steps of constructing the inverted index and processing each key value in the index are relatively efficient.
Consider a key value $k$ which is present in $t$ different rows, out of which $\gamma$ are distinct values that contradict each other. The complexity of \textbf{\texttt{Ver}}'s grouping approach to process the key $k$ in the inverted index takes $O(t)$ running time to identify all contradictions involving $k$. Therefore, this step has complexity linear in the number of contradictions. 



%% file: sections/evaluation_full.tex
\section{Evaluation}
\label{sec:evaluation}

In this section, we answer these research questions:


\begin{myitemize}

\item \textbf{RQ1}: Is \sys{} effective in navigating users to the view that satisfies their requirements? (human problem)

\item \textbf{RQ2}: Is \textsc{View-Distillation} useful for reducing the view choice space and is \textsc{View-Distillation} scalable? (technical problem)

\item \textbf{RQ3}: End-to-end evaluation of \sys{} (technical problem)



\item \textbf{Qualitative Study (QS).} We discuss qualitative differences with QBE systems.

\end{myitemize}

In Appendix.~\ref{appendix:c}, we include several \textbf{Microbenchmarks} where we explore the effect of various query and data parameters on \sys{}'s performance.

\mypar{Datasets and Workload} We use three real-world large-scale datasets in the evaluation. Detailed statistics of these three datasets are shown in Table~\ref{tab:datasets}.

\begin{table}[ht]
\begin{center}
 \begin{tabular}{||c c c c c c||} 
 \hline
 Dataset & \#Tables & \#Columns & \specialcell{$\sim$\# Joinable\\ Columns} & \specialcell{$\sim$ Total\\ \#Rows} & Size\\ [0.5ex] 
 \hline\hline
 ChEMBL & 70 & 446 & 435 & 140M & 6.5GB\\ 
 \hline
 WDC & 10000 & 39939 & 11.6M & 140K & 45MB\\ 
 \hline
 Open Data & 69407 & 2955305 & 28.6M & 900M & 119GB\\
 \hline 
\end{tabular}
 \caption{Characteristics of Datasets}
 \label{tab:datasets}
\end{center}
\end{table}

\begin{myitemize}
    \item \textbf{ChEMBL}: ChEMBL~\cite{ChEMBL} is a database of bioactive molecules with drug-like properties. ChEMBL is large in terms of total data size. However, it has a relatively small number of tables and joinable columns.
    
    \item \textbf{WDC}: WDC is a subset of the web tables corpus~\cite{wdc} containing 10K tables crawled from the web. It has more than 10 million pairs of joinable columns.
    
    \item 
    \textbf{Open Data}~\cite{hu2019viznet}
    This dataset consists of 69K open datasets collected from the Open Data Portal Watch
    \cite{mitlohner2016characteristics,neumaier2016automated}\, which catalogs and monitors 262 open data portals such as NYC Open data, finances.worldbank.org, etc. 
\end{myitemize}

\mypar{System Setup} We ran all experiments on a Ubuntu server with 500GB memory and an Intel(R) Xeon(R) CPU with 48 cores and 2.3GHz speed each. We built \sys{} using python3.6. \sys{} uses Aurum to find join paths without using schema information. 
In \textbf{ChemBL}, we ignore the schema information to simulate pathless scenario and instead use schema to evaluate ground truth.
When searching for join graphs, \sys{} uses by default a maximum of two hops, $\rho = 2$. We set the clustering threshold 
$\theta$ to 1, and the expected number of output views $k$ to be the total number of join graphs so we materialize all join graphs generated from \textsc{Join-Graph-Search}.


\subsection{RQ1: Is \sys{} effective in navigating users to the view that satisfies their requirements?\label{sec:userstudy}}

We conducted a within-subjects user study to answer this question. We give participants a task and expose them to two systems: \textsc{view-presentation} as explained in this paper, and a ranking of views as produced by overlap-based ranking mechanism of \topk{}~\cite{s4}. Their goal is to identify a view that satisfies the task.

\mypar{\textit{Participants}} We recruited 18 students with diverse backgrounds (CS, Economics, Math) from the University. We did not record any personally identifiable information.

\mypar{\textit{Study Procedure}} We design the study to ensure internal validity. Each participant attends a 30-min training session to learn the interface design. During the session, we describe the study, give a tutorial on each interface, and ask participants to solve two randomly-chosen \emph{trial} queries using \sys{} and \topk{}. The goal of the trial task is to familiarize participants with the interfaces. After finishing the trial tasks, each participant solves two different queries with \sys{} and \topk{}, respectively. The order in which the participant uses \sys{} and \topk{} is randomized to avoid ordering and learning effects. Participants work in isolation to avoid biasing each other. After finishing the tasks, participants answer a short survey about their experience with both systems. 




\mypar{\textit{Task Setup}} Participants are exposed to 4 (2 trial, 2 study tasks) of the 5 queries shown in Table~\ref{tab:query_user_study} from WDC~\cite{wdc} dataset. We chose a diverse set of queries involving numerical and textual attributes that generated semantically ambiguous results.

\begin{table}[ht!]
    \centering
    \begin{tabular}{|c|c|c|c|}
    \hline
    Query & Example & \specialcell{\sys{} \\ \#Views} & \specialcell{\topk{} \\ \#Views} \\\hline
       \specialcell{Find views containing\\ IATA code of airports\\in any of these states\\ in the US. }  &  \specialcell{Indiana, Georgia, \\ Virginia, Illinois,\\ Connecticut}  & 397 & 2255 \\ \hline
        \specialcell{Find views containing\\ churches in any of these\\ states in
the US} & \specialcell{Indiana, Georgia, \\Virginia, Illinois,\\
Connecticut} & 397 & 2255 \\ \hline
         \specialcell{Find views containing\\ newspaper companies in\\ any of these cities. } & \specialcell{San Diego,\\Boston,\\ Philadelphia} & 394 & 838\\ \hline
         \specialcell{Find views containing\\population of any of\\ these countries.} & \specialcell{China, Japan, \\United States} & 566 & 2235 \\ \hline
         \specialcell{Find views containing\\ the number of births \\per 1000 population\\in any of these countries.} & \specialcell{China, Japan,\\United States} & 566 & 2235 \\ \hline
    \end{tabular}
    \caption{Tasks used in the user study and \# Views generated by \sys{} / \topk{}.}
    \label{tab:query_user_study}
\end{table}

We manually verified each participant's output and record whether they found a relevant view that answers the query.

\mypar{\textit{Interface Setup}} We setup \sys{}'s \textsc{view-presentation} with different types of question interfaces. The interface asks users if they want to include a specific attribute, collection of attributes or an individual dataset (as discussed in Section~\ref{sec:vpresentation}). We use two different prioritization strategies for each interface: one based on the distance of the question from the input query and other based on the distance of the datasets corresponding to the questions from the input query. We use pre-trained word2vec embeddings to calculate distance. In each interaction, the user can either skip or answer the question or explore the ranking of views to select one view. The scoring model we adopted in \topk{} presents a ranking of views allowing the user to manually explore the options and pick the one that satisfies the input query.


\mypar{\textit{Data Collection and Results}} We log the interactions of each participant with the system for subsequent analysis. We measured interactions and outcomes to answer the following questions (Table~\ref{tab:userresult} presents the study results):





    



\noindent\underline{Q1. Does the user find the relevant view?} $16/18$ participants identify the correct view with \sys, versus only $6$ when using the \topk{} ranking. $12$ participants finished the task without finding any dataset using \topk{} versus only $1$ with \sys{}. The results are statistically significant: we run Fisher's exact test and obtain a p-value of $0.002$. Due to this result, we confirm the sample size is adequate for this study.

\noindent\underline{Q2. Which system would you prefer to search datasets?} $12$ participants prefer searching for datasets with \sys{} and $5$ prefer \topk{} (1 participant was not sure).



\noindent {Q3. If you are to forward the query and the dataset you chose} using \texttt{\textbf{Ver}} same question for \textsc{FastTopK}) to someone else. \underline{How confident are you to share the identified search result} \underline{ for the input query?} $14$ participants were confident with the result they found with \texttt{\textbf{Ver}}. We cannot measure confidence for \textsc{FastTopK} because 12 participants did not find any view.


\noindent\underline{Q4. How difficult is to use \texttt{\textbf{Ver}}}? and {Q5. How difficult is to }  {answer multiple choice questions with \texttt{\textbf{Ver}}?} $14$ participants deemed using \sys{} easy and intuitive and $4$ disagreed. Anecdotally, some participants mentioned that questions asked by \sys{} are easy to answer and do not require in-depth analysis.
Our discussion with the participants revealed that different users preferred different interface designs. For example, some students verified the attribute names before choosing a view while others verified a sample of the records.

\noindent\underline{Time taken} The median participant using \sys{} finds the view within 101 seconds (median) and with a median of $3$ interactions. They take 93 seconds (median) when using \topk{}.


\begin{table}[ht!]
\centering
\begin{tabular}{|c|c|c|}
\hline
\rowcolor{LightCyan}
\multicolumn{3}{|c|}{\specialcell{ Q1. Does the user find a relevant view?}} \\ \hline \hline
& \sys{} & \topk{} \\ \hline
Found & $\textbf{16}^*$ & 6 \\ \hline
Not Found & 2 & 12 \\ \hline 
\multicolumn{3}{|c|}{\specialcell{*Result is statistically significant with  p-value of 0.002}} \\ \hline \hline
\rowcolor{LightCyan}
\multicolumn{3}{|c|}{\specialcell{ Q2. Which system would you prefer to search datasets?}} \\ \hline \hline
\sys{}     &  \topk{} & Unsure \\ \hline
\textbf{12}     &  5 & 1 \\ \hline
\hline
\rowcolor{LightCyan}
\multicolumn{3}{|c|}{\specialcell{ Q3. Confidence in the identified search result}} \\ \hline \hline
& \sys{} & \topk{} \\ \hline
Confident & \textbf{14} & 6 \\ \hline
Not Confident & 4 & 8 \\ \hline
\hline
\rowcolor{LightCyan}
\multicolumn{3}{|c|}{\specialcell{  Q4. How difficult is to use Ver?}} \\ \hline \hline
\multicolumn{2}{|c|}{Intuitive}   & Not Intuitive \\ \hline
\multicolumn{2}{|c|}{\textbf{14}} & 4 \\ \hline \hline
\rowcolor{LightCyan}
\multicolumn{3}{|c|}{\specialcell{ Q5. How difficult is to answer multiple choice questions with Ver?}}            \\ \hline \hline
\multicolumn{2}{|c|}{Easy}      &  Difficult \\ \hline 
\multicolumn{2}{|c|}{\textbf{16}}     &  2 \\ \hline
\end{tabular}
\caption{Summary of survey results. \label{tab:userresult}}
\end{table}

\input{sections/rq2}

\input{sections/4c_scalability}

\input{sections/comparison}

\input{tables/sota-components}

\input{sections/qualitative_qbe}



%% file: sections/rq2.tex
\subsection{RQ2: Does \textsc{View Distillation} reduce the number of views and is \textsc{View-Distillation} scalable?}
\label{subsec:presentation}


We evaluate the effectiveness in reducing result views and the scalability of \textsc{View-Distillation}.


\mypar{Noisy Query Generation} Each query consists of a collection of 2-column, 3-row example values.
To generate the query, we first find a PJ-query that produces a result we call the \emph{ground truth PJ-view}. Columns in the \emph{ground truth PJ-view} are called \emph{ground truth columns}. Then, we generate the $2\times 3$ input queries according to three strategies, \emph{Zero Noise}, \emph{Medium Noise}, and \emph{High Noise}. \emph{Zero noise} means we sample values from the \emph{ground truth columns}. In \emph{Medium noise} we sample $\frac{2}{3}$ values from the \emph{ground truth columns} and $\frac{1}{3}$ from a \emph{noise column}, which is a column with a Jaccard Containment of more than 0.8 with respect to the \emph{ground truth column}. Finally, in \emph{High noise} we sample $\frac{1}{3}$ values from the \emph{ground truth column} and $\frac{2}{3}$ from the \emph{noise columns}.


We generate 5 ground truth queries by sampling join graphs from the ground truth views of ChEMBL and WDC. For each ground truth query, we generate one noisy user query consisting of example values for each of the 3 noise levels. For each noisy user query, we obtain input PJ-views to \textsc{View-Distillation} by getting candidate columns via \textsc{column-selection} component and feeding them to \textsc{Join-Graph-Search} and \textsc{Materializer}.


\input{tables/4c_table}

\subsubsection{\textbf{Compatible and Contained (C1 and C2)}}

Each group of compatible views is reduced to a single view. When views are contained, we keep the larger one. The $C_1$ and $C_2$ columns in Table~\ref{tab:num_views_left} show the number of views left after pruning compatible views and contained views, respectively. 

\mypar{Insights of C1} Around 50\% of candidate PJ-views in Q3 of ChEMBL are compatible because many tables have more than one candidate key. For example, one pair of compatible views are being materialized using the same join tables: \textit{assays} and \textit{cell\_dictionary}, but one view's join key is \textit{cell\_name} and the other's join key is \textit{cell\_description}; since there is a one-to-one mapping between \textit{cell\_name} and \textit{cell\_description}, the views they produced are identical.



\mypar{Insights of C2} Q2 of WDC pruned 18 contained views for zero/medium and 19 for high noise level. The majority of the output views in the output share the same attributes: \textit{State} and \textit{Newspaper Title}. We found that each pair of contained views are joined using different tables but the same join key, \textit{State}. The join key values of one join path are subsumed by the join key values of other join paths, thus the resulting view is contained in another view.   

\subsubsection{\textbf{Complementary (C3)}}

We union complementary views. The complementarity of views depends on their key.
In this experiment, we consider the key that leads to the least reduction (worst case column) and the largest reduction (best key column). Table~\ref{tab:num_views_left} shows that unioning tables reduces the number of views in most queries. In the worst case, tables do not union with each other, as in the case of the WDC  Q3.


\mypar{Insights of C3} The reason WDC Q2 can union many complementary views even in the worst case is that all candidate PJ-views that share the same list of attributes, \textit{State} and \textit{Newspaper Title}, are joined by two tables using the join key \textit{State}; one join table containing the attribute \textit{Newspaper Title} is the same for all the views, while the other containing the attribute \textit{State} is different for each view. The join tables that are different have different coverage of \textit{State} values. Therefore, a lot of candidate PJ-views are complementary based on the candidate key \textit{State} (the worst case). 

For ChEMBL queries, typically one pair of views does not share their contradictory rows with any other pair of views, so no matter which candidate key we choose, it can always lead to unionable complementary views since the contradictory relationships are not transitive across views. For some queries such as Q5, many views do not have valid candidate keys, so there are no unionable views.

\subsubsection{\textbf{Contradictory (C4)}} We construct contradictions from contradictory view pairs by grouping all views that share the same contradiction together. Given a contradiction, we do not have an automated way of choosing a view. However, we calculate the value of pruning by measuring the worst case and best case reduction in the candidate set size.
We sort contradictions in descending order according to their degree of discrimination--the number of views that agree with one side of the contradiction. Then, we select the contradictions sequentially and consider two cases: (a) where the selection leads to the largest reduction of the views  (best-case) and (b) where it leads to the least reduction of the views (worst-case).

\begin{figure}
  \centering
  \includegraphics[width=\linewidth]{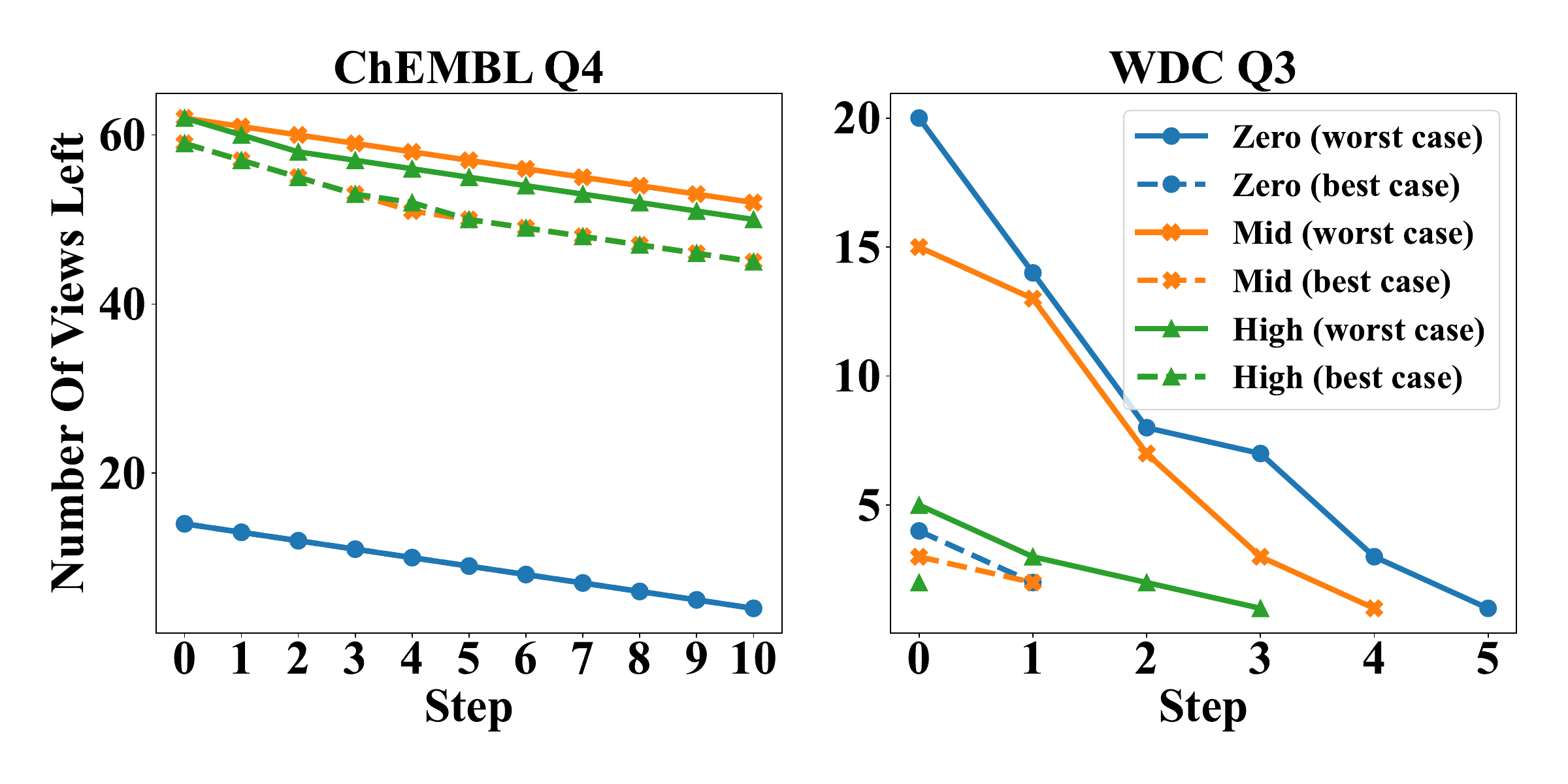}
  \caption{Number of views left at each step after pruning views.}
  \label{fig:num_views_left}
\end{figure}

\F~\ref{fig:num_views_left} shows the number of remaining views after each step (for a maximum of 10 steps) for a selection of queries that present discriminative and non-discriminative contradictions. As expected, when contradictions are not discriminative, the reduction is limited, such as in mid/high noise queries of ChEMBL Q4 in the worst case. However, there are cases where contradictions are quite discriminative and the signal proves effective in reducing the set size, such as in the worst case queries of WDC Q3. 

\mypar{Insights of C4}
In ChEMBL, the output of Q4 mid/high noise queries contains many candidate PJ-views that are joined by varied join tables and keys. The contradictions in candidate views are mainly due to wrong join paths. For example, one view that is partly joined by the two tables \textit{component\_sequences} and \textit{component\_class} on the shared attribute \textit{component\_id}, while the other view is joined by the two tables \textit{component\_sequences} on attribute \textit{description} and \textit{target\_dictionary} on attribute \textit{pref\_name}, when projecting the final attributes, \textit{organism} and \textit{pref\_name}, these two views have contradictions. \textit{component\_id} is a better join key than \textit{description}. However, as we do not know the join information of the input data, we can only perform join operations using all the attributes that are being considered as valid join keys by the discovery engine. 

Moreover, since the contradictions in ChEMBL mainly arise due to different join paths, they typically do not share the same contradictions across multiple views. Each contradictory signal only contains two views. Therefore, the maximum number of views we prune at each step is 1. For WDC queries, however, we are able to prune multiple views in Q3, since the contradictory views have many shared contradictory rows, thus each contradictory signal contains many views. And since we prioritize presenting the more discriminatory contradictions first, we prune multiple views even in the worst case.





%% file: tables/4c_table.tex
\begin{table}[]
\scriptsize
\centering
\begin{center}
{\centering

\begin{tabular}{|c|c|c|c|c|c|c|}
\hline
Query &
  \begin{tabular}[c]{@{}l@{}}Noise\\ level\end{tabular} &
  Original &
  $C_1$ &
  $C_2$ &
  \begin{tabular}[c]{@{}l@{}}$C_3$\\ worst\\ case\end{tabular} &
  \begin{tabular}[c]{@{}l@{}}$C_3$\\ best\\ case\end{tabular} 
  \\ \hline
  
\multirow{3}{*}{\begin{tabular}[c]{@{}c@{}}ChEMBL\\ Q1\end{tabular}} 
                    & Zero &
\ApplyGradientnew{38}{7.0}{36.0}{38.0} & \ApplyGradientnew{36}{7.0}{36.0}{38.0} & \ApplyGradientnew{36}{7.0}{36.0}{38.0} & \ApplyGradientnew{20}{7.0}{36.0}{38.0} & \ApplyGradientnew{20}{7.0}{36.0}{38.0}  \\ \cline{2-7} 
                    & Med  & 
                    \ApplyGradientnew{20}{1.0}{18.0}{20.0} & \ApplyGradientnew{18}{1.0}{18.0}{20.0} & \ApplyGradientnew{18}{1.0}{18.0}{20.0} & \ApplyGradientnew{8}{1.0}{18.0}{20.0}  & \ApplyGradientnew{8}{1.0}{18.0}{20.0}  \\ \cline{2-7} 
                    & High & 
                    \ApplyGradientnew{33}{3.0}{31.0}{33.0} & \ApplyGradientnew{31}{3.0}{31.0}{33.0} & \ApplyGradientnew{31}{3.0}{31.0}{33.0} & \ApplyGradientnew{21}{3.0}{31.0}{33.0}  & \ApplyGradientnew{21}{3.0}{31.0}{33.0}  \\ \hline
                    
\multirow{3}{*}{\begin{tabular}[c]{@{}c@{}}ChEMBL\\ Q2\end{tabular}} 
                    & Zero & 
                    \ApplyGradientnew{59}{7.0}{54.0}{59.0} & \ApplyGradientnew{58}{7.0}{54.0}{59.0} & \ApplyGradientnew{54}{7.0}{54.0}{59.0} & \ApplyGradientnew{51}{7.0}{54.0}{59.0} & \ApplyGradientnew{47}{7.0}{54.0}{59.0}  \\ \cline{2-7} 
                    & Med  & 
                    \ApplyGradientnew{32}{3.0}{30.0}{32.0} & \ApplyGradientnew{32}{3.0}{30.0}{32.0} & \ApplyGradientnew{30}{3.0}{30.0}{32.0} & \ApplyGradientnew{30}{3.0}{30.0}{32.0} & \ApplyGradientnew{29}{3.0}{30.0}{32.0}  \\ \cline{2-7} 
                    & High & 
                    \ApplyGradientnew{41}{3.0}{35.0}{41.0} & \ApplyGradientnew{38}{3.0}{35.0}{41.0} & \ApplyGradientnew{35}{3.0}{35.0}{41.0} & \ApplyGradientnew{32}{3.0}{35.0}{41.0} & \ApplyGradientnew{30}{3.0}{35.0}{41.0}  \\ \hline
\multirow{3}{*}{\begin{tabular}[c]{@{}c@{}}ChEMBL\\ Q3\end{tabular}} 
                    & Zero & 
                    \ApplyGradientnew{58}{4.0}{29.0}{58.0} & \ApplyGradientnew{33}{4.0}{29.0}{58.0} & \ApplyGradientnew{29}{4.0}{29.0}{58.0} & \ApplyGradientnew{23}{4.0}{29.0}{58.0} & \ApplyGradientnew{23}{4.0}{29.0}{58.0}  \\ \cline{2-7} 
                    & Med  & 
                    \ApplyGradientnew{44}{3.0}{17.0}{44.0} & \ApplyGradientnew{21}{3.0}{17.0}{44.0} & \ApplyGradientnew{17}{3.0}{17.0}{44.0} & \ApplyGradientnew{12}{3.0}{17.0}{44.0}  & \ApplyGradientnew{14}{3.0}{17.0}{44.0}  \\ \cline{2-7} 
                    & High & 
                    \ApplyGradientnew{44}{3.0}{17.0}{44.0} & \ApplyGradientnew{21}{3.0}{17.0}{44.0} & \ApplyGradientnew{17}{3.0}{17.0}{44.0} & \ApplyGradientnew{14}{3.0}{17.0}{44.0}  & \ApplyGradientnew{14}{3.0}{17.0}{44.0}  \\ \hline
\multirow{3}{*}{\begin{tabular}[c]{@{}c@{}}ChEMBL\\ Q4\end{tabular}} 
                    & Zero & 
                    \ApplyGradientnew{23}{7.0}{14.0}{23.0} & \ApplyGradientnew{17}{7.0}{14.0}{23.0} & \ApplyGradientnew{14}{7.0}{14.0}{23.0} & \ApplyGradientnew{14}{7.0}{14.0}{23.0} & \ApplyGradientnew{14}{7.0}{14.0}{23.0}  \\ \cline{2-7} 
                    & Med  & 
                    \ApplyGradientnew{83}{8.0}{69.0}{83.0} & \ApplyGradientnew{74}{8.0}{69.0}{83.0} & \ApplyGradientnew{68}{8.0}{69.0}{83.0} & \ApplyGradientnew{62}{8.0}{69.0}{83.0} & \ApplyGradientnew{59}{8.0}{69.0}{83.0}  \\ \cline{2-7} 
                    & High & 
                    \ApplyGradientnew{83}{8.0}{69.0}{83.0} & \ApplyGradientnew{74}{8.0}{69.0}{83.0} & \ApplyGradientnew{68}{8.0}{69.0}{83.0} & \ApplyGradientnew{62}{8.0}{69.0}{83.0} & \ApplyGradientnew{59}{8.0}{69.0}{83.0}  \\ \hline
\multirow{3}{*}{\begin{tabular}[c]{@{}c@{}}ChEMBL\\ Q5\end{tabular}} 
                    & Zero & 
                    \ApplyGradientnew{24}{8.0}{15.0}{24.0} & \ApplyGradientnew{18}{8.0}{15.0}{24.0} & \ApplyGradientnew{15}{8.0}{15.0}{24.0} & \ApplyGradientnew{15}{8.0}{15.0}{24.0} & \ApplyGradientnew{15}{8.0}{15.0}{24.0}  \\ \cline{2-7} 
                    & Med  & 
                    \ApplyGradientnew{64}{8.0}{52.0}{64.0} & \ApplyGradientnew{57}{8.0}{52.0}{64.0} & \ApplyGradientnew{51}{8.0}{52.0}{64.0} & \ApplyGradientnew{46}{8.0}{52.0}{64.0} & \ApplyGradientnew{46}{8.0}{52.0}{64.0}  \\ \cline{2-7} 
                    & High & 
                    \ApplyGradientnew{33}{13.0}{20.0}{33.0} & \ApplyGradientnew{23}{13.0}{20.0}{33.0} & \ApplyGradientnew{20}{13.0}{20.0}{33.0} & \ApplyGradientnew{20}{13.0}{20.0}{33.0} & \ApplyGradientnew{20}{13.0}{20.0}{33.0} \\ \hline
\multirow{3}{*}{\begin{tabular}[c]{@{}c@{}}WDC\\ Q2\end{tabular}} 
                    & Zero & 
                    \ApplyGradientnew{44}{4.0}{25.0}{44.0} & \ApplyGradientnew{39}{4.0}{25.0}{44.0} & \ApplyGradientnew{21}{4.0}{25.0}{44.0} & \ApplyGradientnew{8}{4.0}{25.0}{44.0}  & \ApplyGradientnew{6}{4.0}{25.0}{44.0} \\ \cline{2-7} 
                    & Med  &  
                    \ApplyGradientnew{42}{3.0}{23.0}{42.0} & \ApplyGradientnew{37}{3.0}{23.0}{42.0} & \ApplyGradientnew{19}{3.0}{23.0}{42.0} & \ApplyGradientnew{5}{3.0}{23.0}{42.0}  & \ApplyGradientnew{3}{3.0}{23.0}{42.0} \\ \cline{2-7} 
                    & High & 
                    \ApplyGradientnew{39}{5.0}{20.0}{39.0} &
                    \ApplyGradientnew{34}{5.0}{20.0}{39.0} &
                    \ApplyGradientnew{15}{5.0}{20.0}{39.0} &
                    \ApplyGradientnew{6}{5.0}{20.0}{39.0} &
                    \ApplyGradientnew{5}{5.0}{20.0}{39.0}
                     \\ \hline
\multirow{2}{*}{\begin{tabular}[c]{@{}c@{}}WDC\\ Q3\end{tabular}} 
                    & Zero & 
                    \ApplyGradientnew{20}{1.0}{20.0}{20.0} & \ApplyGradientnew{20}{1.0}{20.0}{20.0} & \ApplyGradientnew{20}{1.0}{20.0}{20.0} & \ApplyGradientnew{20}{1.0}{20.0}{20.0}  & \ApplyGradientnew{4}{1.0}{20.0}{20.0} \\ \cline{2-7} 
                    & Med  & 
                    \ApplyGradientnew{15}{1.0}{15.0}{15.0} & \ApplyGradientnew{15}{1.0}{15.0}{15.0} & \ApplyGradientnew{15}{1.0}{15.0}{15.0} & \ApplyGradientnew{15}{1.0}{15.0}{15.0}  & \ApplyGradientnew{3}{1.0}{15.0}{15.0} \\ \hline                    
\end{tabular}
}
\end{center}
\caption{Effect of view distillation based on 4C signals on number of view. We excluded queries that have less than 10 original number of views.}
\label{tab:num_views_left}
\end{table}

%% file: sections/4c_scalability.tex
\subsubsection{Scalability}
\label{subsec:4c_scalability}

\begin{figure}
  \centering
  \includegraphics[width=\linewidth]{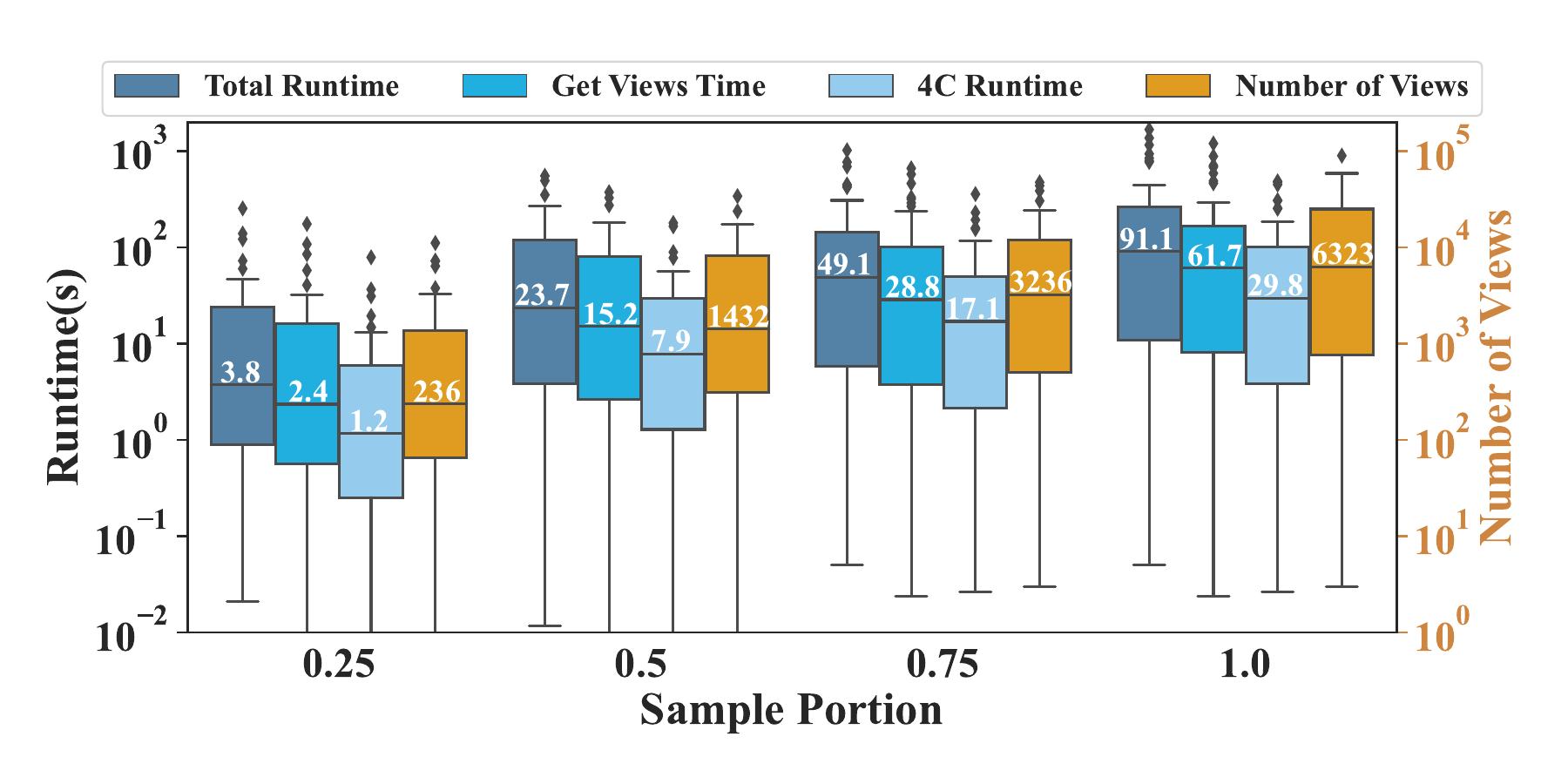}
  \caption{Total Runtime, Get Views Time, and 4C Runtime of \textsc{View-Distillation} for each sample portion on left y-axis, and number of views on the right y-axis.}
  \label{fig:scalability1}
\end{figure}


\begin{figure}
    \centering
    \subfloat[4C Runtime for different steps]{{\includegraphics[width=0.5\columnwidth]{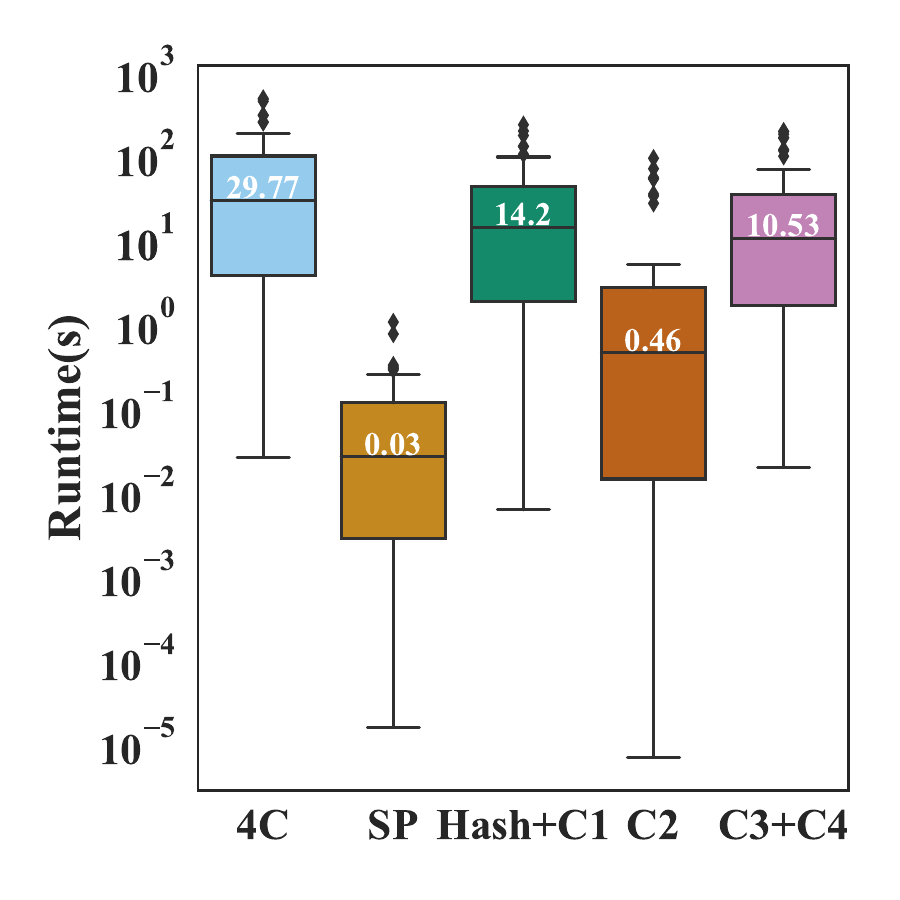} }}
    \subfloat[Total Runtime of Ver over 50 queries]{{\includegraphics[width=0.5\columnwidth]{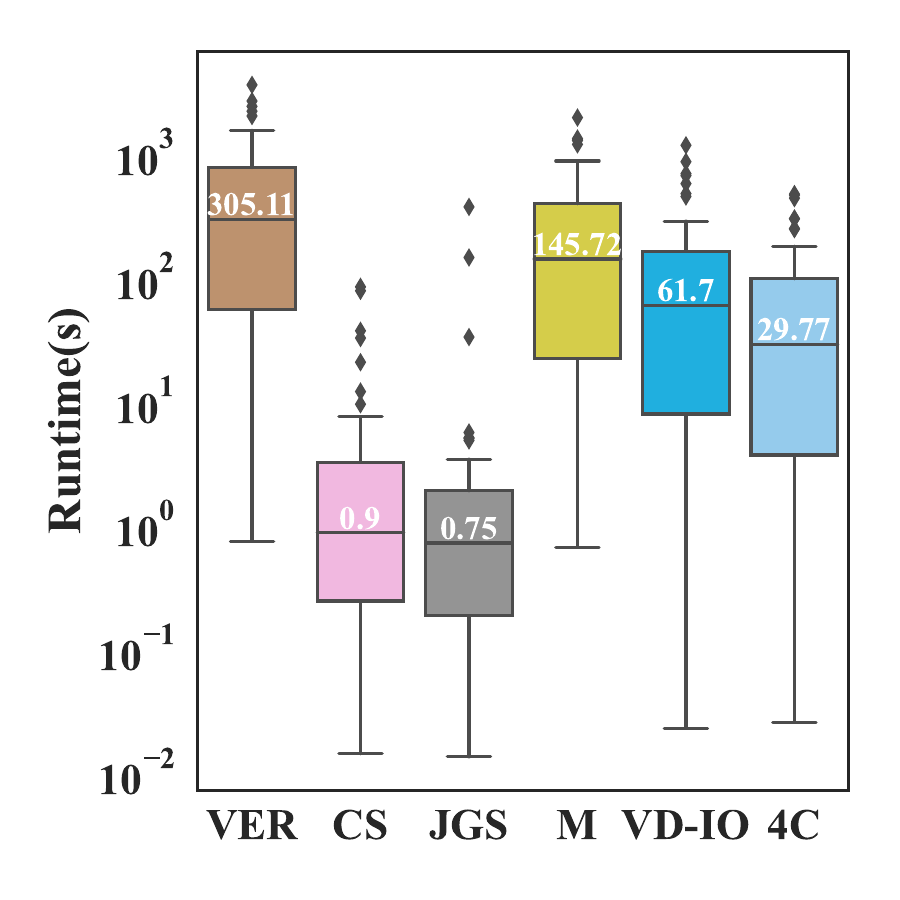} }}
    \caption{(a) Time to execute different steps in 4C for sample portion=1 (SP=Schema Partition). (b) CS=\textsc{Column-Selection}, JGS=\textsc{Join-Graph-Search}, M=\textsc{Materializer}, VD-IO=Get Views Time in \textsc{View-Distillation}, 4C=4C Runtime in \textsc{View-Distillation}.}
    \label{fig:scalability2}
\end{figure}

In this experiment, we evaluate the scalability of \textsc{View-Distillation} using $50$ randomly sampled queries from the OpenData dataset along with 3 datasets built using random samples of $25\%, 50\%,$ and $75\%$ tables from the original\footnote{The subsampling was performed to ensure that all datasets present in a smaller size version are also present in the larger sample.}. Figure~\ref{fig:scalability1} shows the distribution of the number of views for each subsample (see 2nd y-axis). Note that we use boxplots to report the min, max, 25th, median, 75th, and max runtimes given the varied complexity of queries. The total runtime (y-axis) grows linearly with the number of views. Concretely, and as a proxy summary statistic, we observe that \texttt{\textbf{Ver}} takes $1.16$ seconds to calculate 4C categories for around $236$ views (median for $25\%$ sample size) and around $30$ seconds for $6500$ views (median of $100\%$ sample size); we observe similar growth when comparing other percentiles. The algorithm's scalability is limited by the time to read views from disk, which grows with the total amount of data. The total `4C Runtime' is small in comparison.


Figure~\ref{fig:scalability2}(a) zooms into the time taken by 4C (`4C Runtime' in Figure~\ref{fig:scalability1}) to understand the impact of different parts of Algorithm~\ref{alg:4c_main}. The schema group classification and identification of contained views are efficient and require less than $0.5$ seconds. Hashing dominates runtime as it needs to hash each row in the set of candidate views. The time to find contradictory or complementary views involves two main steps: i) find candidate keys for each view, which is linear in the number of total rows; ii) find the actual contradictions based on the inverted index, which is linear in the number of contradictions. When the number of contradictions is small the time to find candidate keys dominates.

\mypar{Effectiveness of View Distillation}
In this experiment, we evaluated the reduction ratio (fraction of view pruned) of merging compatible and contained views for the $100\%$ sample size. We observed that $50\%$ of the queries had a reduction ratio of more than $17.5\%$, while $25\%$ had a reduction ratio of $63\%$. This demonstrates the effectiveness of \textsc{view-distillation} to efficiently prune contained and compatible views.

%% file: sections/comparison.tex
\subsection{End-to-End Evaluation}
\label{subsec:end2end_time}

We first present an end-to-end experiment of Ver using different implementations of \textsc{View-Specification}. We then study the QBE implementation in more detail to understand the effect of different baselines for different components.

\subsubsection{Alternative View Specification Implementations}
In this experiment, we implement 3 view specification methods i) QBE (Ver's default); ii) Keyword search; and iii) Attribute search. We use $10$ randomly chosen queries from OpenData. Every query runs within $11$ minutes (for $\approx 27$K views) with QBE interface, $13$ minutes ($\approx 500$ views) for keyword interface and $30$ minutes ($\approx 1000$ views) for attribute interface. The views generated by keyword and attribute interfaces contain a large number of columns as compared to QBE, contributing to higher running time for these implementations. We further run \textsc{View-distillation} to merge compatible and contained views, followed by \textsc{View-presentation} with a simulated user. We simulated the user to answer questions correctly. We observe that the user identified the ground truth view in as few as $20$ queries for around $500$ views and less than $100$ queries for $3000$ views.  This evaluation demonstrates the effectiveness of our \textsc{view-distillation} and \textsc{presentation }to effectively prune the search space and help the user identify relevant views. In terms of runtime, \textsc{view-presentation} produces questions for users in less than $10^{-3}$ seconds.

We now dive deeper into the QBE implementation to understand the intricacies of our implementation.
\subsubsection{Runtime Comparison}
In this experiment, we report the distribution of runtimes for the sample of $50$ queries used to evaluate 4C's scalability.

Figure~\ref{fig:scalability2} (b) shows that the total runtime is below 305.11 seconds for $50\%$ of the queries. The bottleneck is the \textsc{materializer} and the time taken to read views from the disk, which require $145$ and $62$ seconds for $50\%$ of the queries, respectively. \textsc{materializer}'s runtime is linear with the number of join graphs generated for a query. We use pandas library in Python to materialize the join and read the view from disk, which could be optimized by using a database. 

The median runtime of \textsc{Column-Selection} and \textsc{Join-Graph-Search} is less than 1 second. There are fewer than $3$ outlier queries which take more than $100$ seconds for \textsc{Column-Selection} because these queries are too general (numerical values without semantic meaning) and the input query is present in more than $100K$ datasets. We do not report the time taken by \textsc{view-presentation} as it requires human-in-the-loop. However, we observe that the median time taken to initialize the component is $73$ seconds but it takes less than $0.5$ millisecond per question. Low latency of asking a question helps to ensure interactive performance of \texttt{\textbf{Ver}}.


\input{sections/rq3}

%% file: sections/rq3.tex
\subsubsection{RQ3: Column Selection and Join Graph Search}
\label{subsec:specandsearch}

In this section we ask: do \textsc{column-selection} and \textsc{join-graph-search} find relevant PJ-views given a noisy input query over pathless table collections?

\mypar{Workload} We generate noisy user queries as described in Section~\ref{subsec:presentation}. We generate 5 noisy user queries consisting of example values for each ground truth view and for each of the 3 noise levels. This results in a total of 150 noisy user queries across both datasets and noise levels.

\mypar{Baselines} We compare the \textsc{column-selection} component in \sys{} with two other baselines:

$\bullet$~\textsc{Select-All.} This baseline (implemented from \textit{\topk{}}~\cite{s4}) selects any column that contains at least an example from the input query.

$\bullet$~\textsc{Select-Best.} This baseline selects the column that contains the highest number of examples from the input query. This is the selection strategy implemented by \textit{SQuID}~\cite{squid}.

After running the baselines, we feed the returned candidate columns to the \textsc{join-graph-search} component that finds joinable groups of tables identified in the \textbf{Join Graph Enumeration} stage, identifies join graphs, and materializes them to produce the set of candidate PJ-views.

\begin{table}[]
    \centering
    \begin{tabular}{|c|c|c|c|c|c|c|c|c|}
    \hline
   
        \multicolumn{9}{|c|}{Ground Truth Hit Ratio}  \\ \hline \hline
        
        \multicolumn{3}{|c|}{Zero Noise} & \multicolumn{3}{c}{Mid Noise} &
        \multicolumn{3}{|c|}{High Noise} \\ \hline
        SA & SB & CS & SA & SB & CS & SA & SB & CS \\ \hline
        1.0 & 1.0 & 1.0 & 1.0 & 0.08 & 1.0 & 1.0 & 0.02 & 0.96 \\ \hline 
    \end{tabular}
    \caption{Ground truth hit ratio over 150 queries in input
workload split by noise level in the input query (SA: Select-All, SB: Select-Best, CS: Column-Selection).}
    \label{tab:agg_hits}
\end{table}


\begin{figure}
  \centering
  \includegraphics[width=\linewidth]{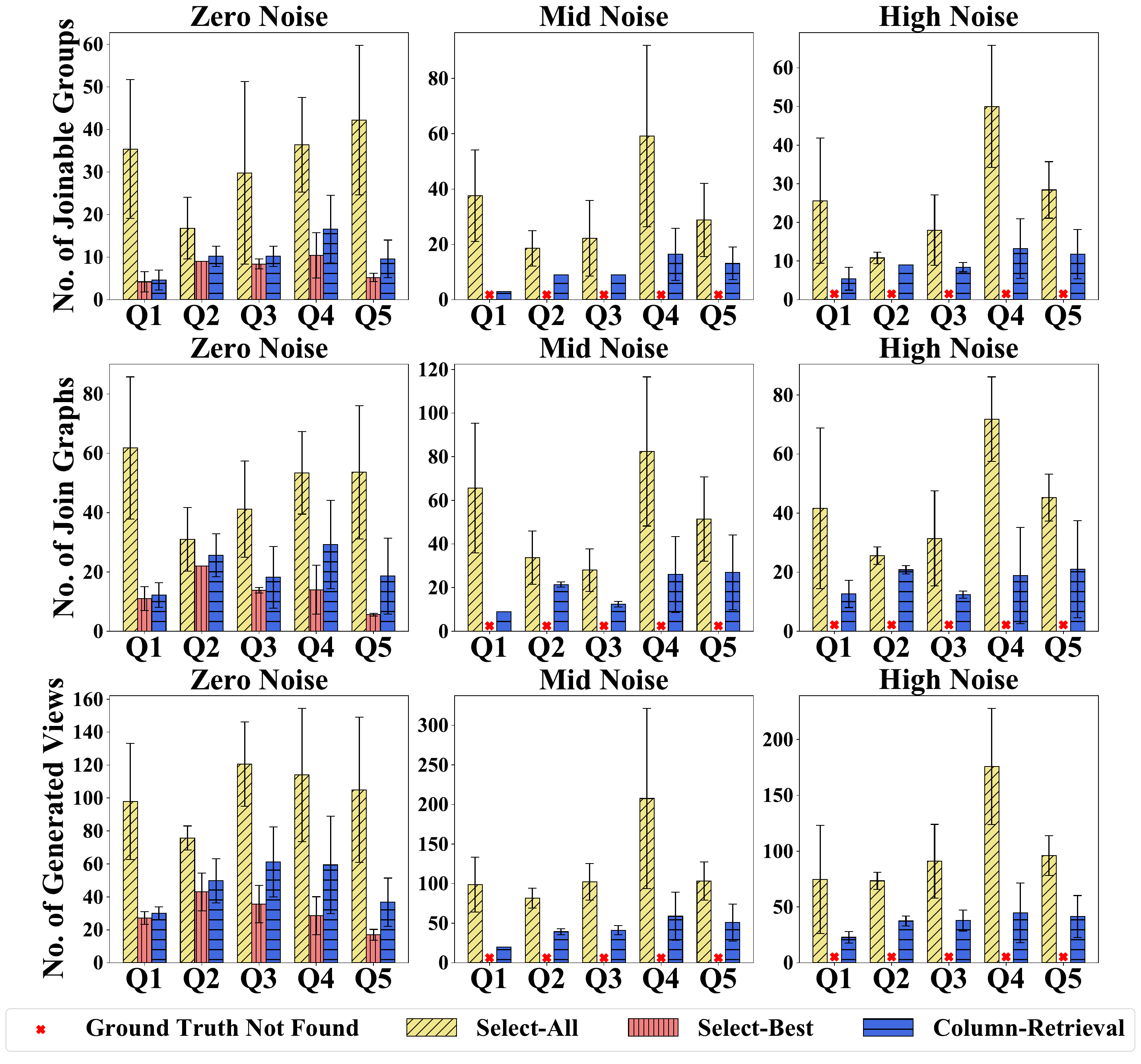}
  \caption{\#joinable groups, join graphs and views on ChEMBL}
  \label{fig:chembl_view_search_info}
\end{figure}

\mypar{Experiment and Metrics} We obtain the output set of candidate PJ-views for each of the 150 input user queries in the workload as described in \textbf{Noisy Workload Generation}. We consider two metrics. First, whether the \emph{ground truth view} is part of the candidate PJ-views, \ie the system finds the required view. In particular, we measure the \emph{Ground Truth Hit Ratio} that determines the ratio of input queries for which the system finds the \emph{ground truth view}. Second, we measure the size of the set of candidate PJ-views. Given two systems that find the \emph{ground truth view}, we prefer the one producing the smaller set of candidate PJ-views. Smaller candidate PJ-views indicate lower runtime (as we will demonstrate later) and more importantly, facilitates the job of the \textsc{view-presentation} stage, \eg consider a human who needs to look at each view in the candidate set to select the right one.

\mypar{Results} Table~\ref{tab:agg_hits} shows the \emph{Ground Truth Hit Ratio} across the 150 queries, for each baseline, and grouped by different input query noise levels in the \textsc{X}-axis. When there is no noise in the input query, then all baselines perform well and find the \emph{ground truth view}. As the noise in the input query increases, the \textsc{Select-Best} strategy crumbles because of its over-reliance on columns that contain all values in the input query. This demonstrates that when input queries contain noise, the \textsc{Select-Best} strategy is inadequate.

\begin{figure}
  \centering
  \includegraphics[width=\linewidth]{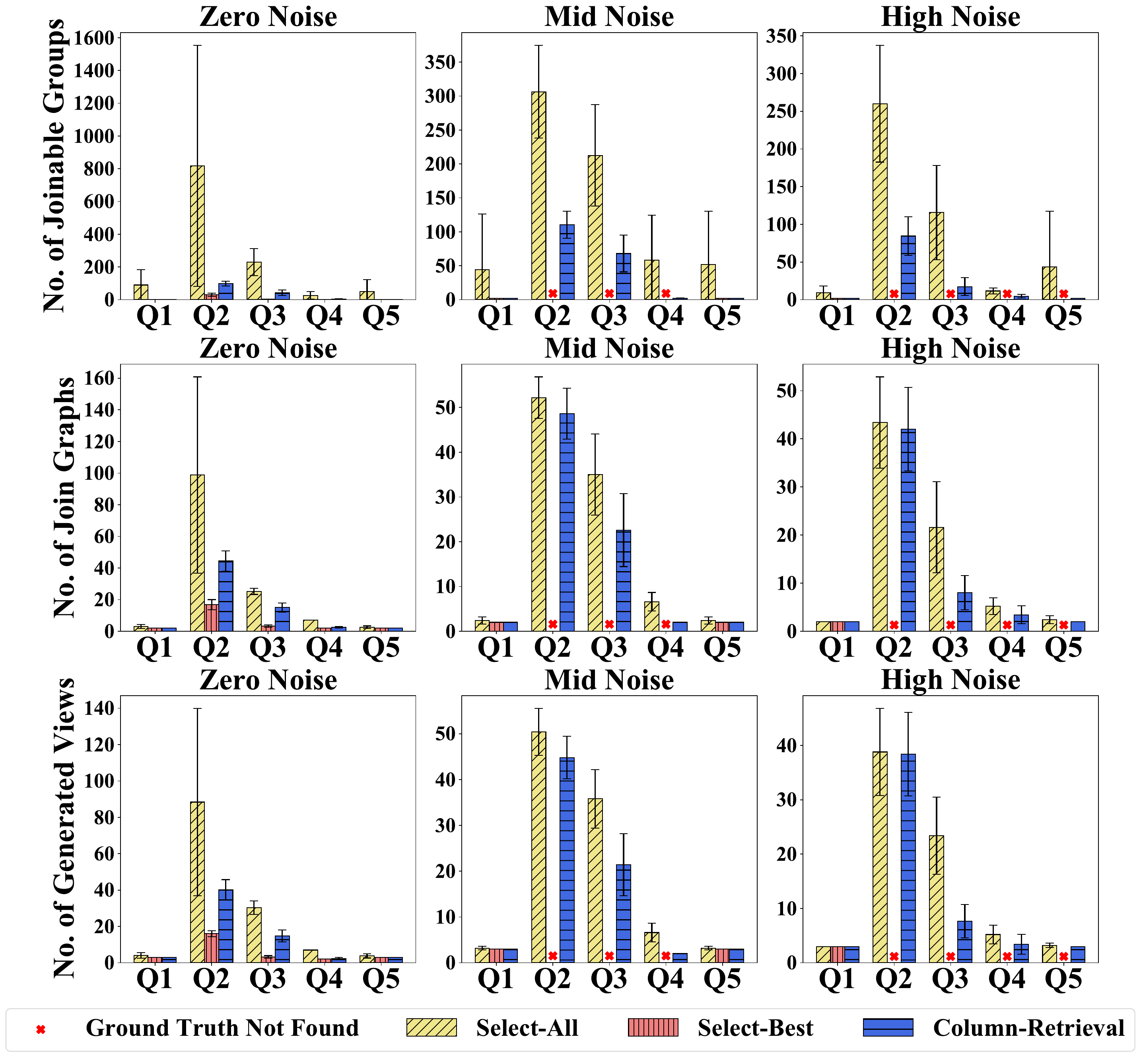}
  \caption{\#joinable groups, join graphs and views on WDC}
  \label{fig:wdc_view_search_info}
\end{figure}

With both \textsc{Select-All} and \textsc{column-selection} consistently finding the \emph{ground truth view}, the next question is at what cost. \F~\ref{fig:chembl_view_search_info} and \F~\ref{fig:wdc_view_search_info} show results for ChEMBL and WDC respectively. Each $3\times 3$ grid shows the size of the set of candidate PJ-views at the top for the three noise levels. The results clearly indicate that for all queries across both datasets and noise levels, the set of the candidate PJ-views is always significantly larger in the case of \textsc{Select-All} than in the case of \textsc{column-selection}. Since both baselines find the \emph{ground truth view}, the smaller sets are preferred.

\begin{figure}
  \centering
  \includegraphics[width=\linewidth]{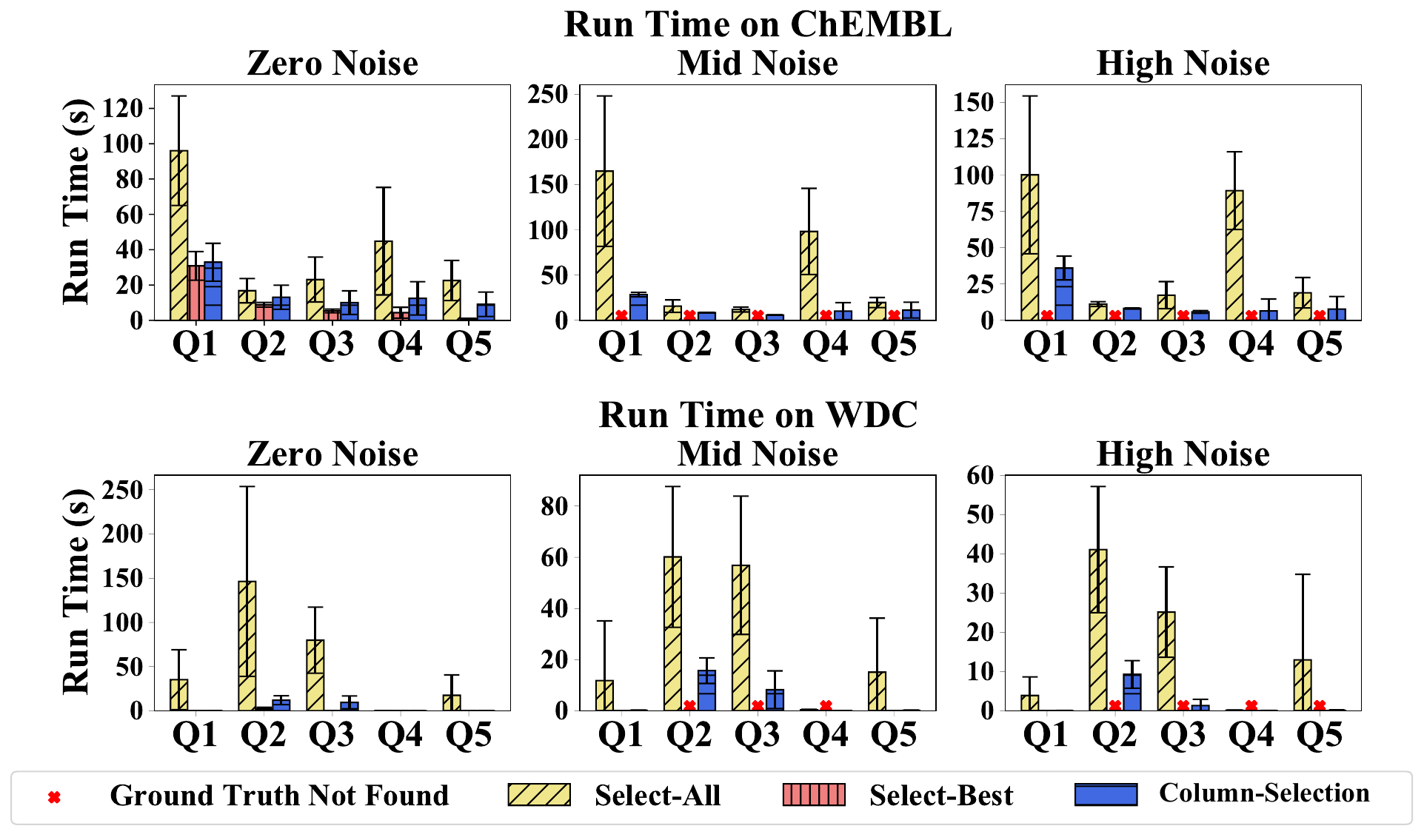}
  \caption{Run time of \textsc{column-selection} + \textsc{join-graph-search} + \textsc{Materializer} on ChEMBL and WDC}
  \label{fig:vs_run_time}
\end{figure}

The \textsc{Select-All} retrieval strategy selects many columns and produces much larger joinable groups than necessary (see \textit{No. of Joinable Groups} in \F~\ref{fig:chembl_view_search_info} and \F~\ref{fig:wdc_view_search_info}). Larger joinable groups, in turn, lead to a larger number of join graphs--sometimes up to $4\times$ more, see \textit{No. of Join Graphs} in \F~\ref{fig:chembl_view_search_info} and \F~\ref{fig:wdc_view_search_info})--which results in more candidate PJ-views. \F\ref{fig:vs_run_time} shows that larger sets of candidate PJ-views leads to higher runtime. The runtime of \textsc{column-selection} + \textsc{join-graph-search} is an order of magnitude lower than the \textsc{Select-All} strategy.

%% file: tables/sota-components.tex
\begin{table*}[h!]
\scriptsize
    \centering
    \begin{tabular}{|c|c|c|c|c|c|c|c|c|}
    \hline 
     & \textbf{Technique}   & \multicolumn{2}{c|}{\textbf{\specialcell{View\\ Specification}}} & \textbf{\specialcell{Column\\ Selection}} &  \textbf{\specialcell{Discovery \\ Engine}}  & \textbf{\specialcell{Join Graph \\ Search}}  & \textbf{\specialcell{View\\ Distillation}} & \textbf{\specialcell{View\\ Presentation}} \\\hline
     &&Input Type & Handles Noise & &Require PK/FK&&&\\
       \hline\hline
       \multirow{4}{*}{\rotatebox[origin=c]{90}{\textbf{QBE}}} & SQuID \cite{squid}  & Relational& N &  Automatic  & Y  & \specialcell{Online} & N & N\\
       \cline{2-9}
      
       & S4: FastTopK \cite{s4} & Relational& Y &  Automatic & Y   & Online & Individual signal & N \\
        \cline{2-9}
        & MWeaver \cite{mweaver}  & Relational & N & Automatic & Y & Online & Individual signal& N \\
        \cline{2-9}
         & DuoQuest \cite{duoquest}  & Natural language & Y &  Automatic & Y & Online & Individual signal & N \\
        
       \hline
       \multirow{4}{*}{\rotatebox[origin=c]{90}{\textbf{QRE}}}   
       
       & TALOS \cite{qbo} &  \multicolumn{2}{c|}{N} & Automatic & Y & Online & Individual signal & N \\
       \cline{2-9}
       & PALEO-J \cite{paleo-j} & \specialcell{Ranks tuples} & N & Automatic & Y & Online & Individual signal & N \\
      \cline{2-9}
      & SQLSynthesizer \cite{sql-synthesizer} &  Relational & N & Automatic & Y & Online & Individual signal & N \\
      \cline{2-9}
       & REGAL+ \cite{regal-plus} &  \multicolumn{2}{c|}{N} & Automatic & Y & Online & N & N \\\cline{2-8}
       \hline
       \multirow{5}{*}{\rotatebox[origin=c]{90}{\textbf{Data Disc.}}}  &Aurum~\cite{aurum} & \multicolumn{2}{c|}{N} & N & N, \specialcell{Offline index} & Online & N & N\\
       \cline{2-9}
         &Josie~\cite{zhu2019josie} & \multicolumn{2}{c|}{N} & N & N, \specialcell{Offline index} & Online & N & N\\
         \cline{2-9}
         &TableUnion~\cite{tableUnionSearch} & \multicolumn{2}{c|}{N} & N & N, \specialcell{Offline index} & Online & N & N\\\cline{2-9}
         &Lazo~\cite{fernandez2019lazo} & \multicolumn{2}{c|}{N} & N & N, \specialcell{Offline index} & Online & N & N\\\cline{2-9}
         &LSHEnsemble~\cite{zhu2016lsh} & \multicolumn{2}{c|}{N} & N & N, \specialcell{Offline index} & Online & N & N\\\cline{2-9}
         &PEXESO~\cite{dong2021efficient} & \multicolumn{2}{c|}{N} & N & N, \specialcell{Offline index} & Online & N & N\\\hline
  \multirow{3}{*}{\rotatebox[origin=c]{90}{{\textbf{VP}}}}  &Voyager~\cite{wongsuphasawat2017voyager} &  \multicolumn{2}{c|}{N} & N & N& N & N & Y\\
  \cline{2-9}
     &SeeDB~\cite{vartak2014seedb} &  \multicolumn{2}{c|}{N} & N & N& N & N & Y\\
    \cline{2-9}
    &NorthStar~\cite{kraska2018northstar}&  \multicolumn{2}{c|}{N} & N & N& N & N & Y\\
    \cline{2-9}
     &RONIN~\cite{ronin}&  \multicolumn{2}{c|}{N} & N & N& N & N & Y\\
     \hline
       & \textbf{\sys{}} &  Relational & Y & \specialcell{Automatic,\\ Interactive} & \specialcell{N, Offline index} & Online & \specialcell{Individual,\\ Dependent signal} & Y \\
       \hline
    \end{tabular}
    \caption{Overview of SOTA. VP = View Presentation and Data Disc. = Data Discovery. N means the system does not implement the component. Individual signals means the approach only computes a statistic over candidate views. Dependent signals consider dependence between candidate views like 4C 
    signals.  }
    
    \label{tab:components}
\end{table*}

%% file: sections/qualitative_qbe.tex
\subsection{QS: Analysis of existing QBE systems}
\label{subsec:comparisons}


\noindent\textbf{SQuID}~\cite{squid} does not scale to pathless table scenarios. \textit{SQuID} precomputes an abduction-ready database ($\alpha$DB) that requires human input to select the pairs of table key and attributes of interest from other tables. Without human input the number of combinations grows large, especially given that in pathless table collections some join path   information will be wrong. Given that the size of $\alpha$DB can be as large as the original table---for example, \emph{component\_sequences} of ChEMBL is 5.9M and results in an $\alpha$DB size of 8.1M---the storage footprint would multiply.
Therefore, we were only able to test \textit{SQuID} on a dataset containing a handful of tables. Without a deep understanding of the input dataset---which we lack in pathless table collections---we can only provide limited information to the system to compute the $\alpha$DB, thus resulting in poor query performance. A more reasonable use of \textit{SQuID} is to employ it as a downstream component to \sys{}, where the input is a narrowed-down version of candidate PJ-views.

\noindent\textbf{Duoquest}~\cite{duoquest} lets users specify a natural language query (NLQ) along an input table containing example tuples and additional information about the attributes (Table Sketch Query). It outputs candidate SQL queries ranked from highest to lowest confidence based on the user's input queries. Navigating that ranking presents challenges to the user in pathless table scenarios because the candidate query may consist of incorrect join tables or keys due to noise in data. An additional presentation phase is necessary to navigate the user among the candidate queries. A more reasonable use of \textit{Duoquest} is to employ it as an implementation of the \textsc{view-specification} component, giving users alternative ways of describing their queries.



%% file: sections/related_work.tex
\section{Related Work}
\label{sec:relatedwork}




We use Table~\ref{tab:components} to overview the related work.

\mypar{View Specification} Common specification interfaces are keywords~\cite{auctus}, APIs, table search~\cite{zhang2019juneau}, and QBE (relational in the table). Each of these interfaces caters to different discovery needs and can be plugged into our reference architecture. Although we implemented QBE in \textbf{\texttt{Ver}}, we evaluated the performance of alternative view specification implementations such as keywords, and attribute-based search interface.

\mypar{Data Discovery} Some discovery engines~\cite{aurum, d3l, zhang2019juneau, tableUnionSearch, elexamples} identify join paths in repositories and their emphasis is often effectiveness and scalability. There are many other techniques that identify join paths using techniques such as summaries~\cite{santos2021correlation}, etc. Most discovery engines leverage key detection techniques as a building block to identify join paths~\cite{key1, key2}. Complementary work to identifying join paths is on automatically detecting transformations to expand the set of joinable columns~\cite{transform1, transform2, transform3}. Any improvement to the detection of join paths can be incorporated into the reference architecture presented here. The `Discovery Engine' column shows what techniques depend on existing join paths and how the navigation is done (column `Join Graph Search').


\mypar{View Distillation} Ranking is a well-explored topic in data management and many approaches naturally leverage this technique to sort candidate results, see column `View Distillation' in the table, where Individual Signal refers to the ranking score. Orthogonally, view summarization~\cite{joglekar2017interactive}, and automated data exploration techniques concentrate in sorting through data and offer an alternative way of navigating the data, like view distillation. Unlike distillation, none of these techniques leverage 4C categories.

\mypar{View Presentation} There is related work on designing effective visualizations to assist users in exploring pathless data collections~\cite{hu2019viznet,bikakis2016exploration, ronin} and view recommendations~\cite{zhang2019viewseeker,zhang2020interactive}. These techniques can be useful to implement the human components of \sys{}. Table~\ref{tab:components} summarizes some techniques.

\mypar{Applying QBE to Pathless Table Collections}
We present the prior QBE systems in Table~\ref{tab:components}. Other QBE techniques~\cite{mweaver,shen14,s4,bonifati,qbo,view_definition,sql-synthesizer,complex_join_queries,hao,paleo-j,scythe,REGAL,fastqre} are not a system, are not designed to handle pathless scenarios and do not focus on  \textbf{Challenges 4 and 5}.





\textit{Supports Pathless?}  Existing QBE \cite{squid,duoquest,s4,shen14,mweaver} and Query Reverse Engineering (QRE) systems \cite{hao, paleo, qbo,view_definition} are designed for databases with well-defined path information, i.e., primary key/foreign key relationships, and they will generate spurious results when executed over noisy join paths. Bonifati et al. \cite{bonifati} learns join predicates without assuming the existence of join paths. But the approach requires performing a Cartesian product on relevant tables, which introduces a scalability challenge even in moderate size databases. Other techniques that use QBE interface, either assume a different data format, such as knowledge bases~\cite{kgqbe1}, or consider a different view specification interface, such as exemplar queries in \cite{exemplar, exemplar2, exemplar3,pimplikar2012answering} that define a more general notion of queries than query-by-example.

%% file: sections/conclusions.tex
\section{Conclusions}
\label{sec:conclusions}

We presented a reference architecture for the discovery of PJ-views over pathless table collections, and a system, \sys{}, that addresses both technical and human problems of view discovery. \sys{} efficiently addresses the challenges of large-scale pathless table collections by combining different components, including \textsc{view-distillation} and \textsc{view-presentation}. 


%% file: sections/acknowledgement.tex
\section{Acknowledgement}
\label{sec:acknowledgement}

We thank the Chameleon cloud platform~\cite{keahey2020lessons} for providing the resources necessary to conduct this research. We thank all participants of our user study and the anonymous reviewers for helping us improve the work. This work was partially supported by the NSF Convergence Accelerator (Award Number \#2040718) and NSF grant \#2030859.

%% file: sections/appendix.tex
\appendix
\label{sec:appendix}





\subsection{Discovery Index Construction}
\label{appendix:a}

\sys{} leverages Aurum~\cite{aurum} to efficiently build a discovery index over large collections of data. \sys{} uses the following functions provided by Aurum:


\mypar{\textsc{Search-Keyword(target, fuzzy)}} Given an input string, it returns columns that contain the string in either the attribute name, or in the values, as specified by \textsc{target}. The match can be exact or fuzzy, after specifying a maximum Levenshtein distance.

\mypar{\textsc{Neighbors(threshold)}} Given an input column, it returns all neighbors with a Jaccard containment~\cite{fernandez2019lazo} above the input \textsc{threshold}.

\mypar{\textsc{Generate-Join-Graphs(tables, $\rho$)}} Given a set of tables, it returns all join graphs connecting input tables, via inclusion dependencies. Each edge in the join graph has a maximum number of hops, $\rho$.

\subsection{Column Selection}
\label{appendix:b}

\textsc{Column-Selection} generates a set of columns $\textsc{Cand}(A_i)$ given user-provided input examples, $\chi.A_i$. To deal with noisy inputs, the component clusters candidate columns and returns clusters with top-$\theta$ scores. Setting $\theta=1$ returns clusters with the highest overlap with $\chi.A_i$ (there could be ties). In contrast, $\theta=\infty$ returns any column with non-empty overlap. This relaxed design makes the component more robust to noisy inputs than methods based on \emph{exact containment}~\cite{squid,duoquest}, and more efficient than methods based on \emph{non-empty containment}.

\IncMargin{1em}
\begin{algorithm}\SetKwFunction{Union}{Union}
\scriptsize
\SetKwInOut{Input}{Input}\SetKwInOut{Output}{Output}
	\Input{Example value set $\chi.A_i$ for a column $A_i$, Discovery Index $I$, Clustering threshold $\theta$  } 
	\Output{$\textsc{Cand}(A_i)$ Candidate columns and scores.}
	 \BlankLine 
	 
	 $\textsc{Cand}(A_i) \leftarrow \phi$\\
	 	 \For{ $e\in \chi.A_i$ \label{crline2}}
	 {
	    
	    $\textsc{Cand}(e) \leftarrow \textsc{Search-Keyword}(e)$\\
	 	$\textsc{Cand}(A_i) \leftarrow \textsc{Cand}(A_i) \cup \textsc{Cand}(e)$\label{crline4}
	 }
	 
	 $ \mathcal{C} \leftarrow \textsc{Connected-Component}(\textsc{Cand}(A_i), I)$\label{crline5}\\ 
	 \For{ $Cluster\in \mathcal{C}$ }{
	    $\textsc{Score}(Cluster) = \max_{col\in Cluster} (| col \cap \chi.A_i |) \label{crline7} $
	 }
	 $\mathcal{C}'\leftarrow$ top-$\theta$ clusters in $\mathcal{C}$ based on $\textsc{Score}$\label{crline8}\\
	 $\textsc{Cand}(A_i) \leftarrow \bigcup_{T\in \mathcal{C}'} T$\\
	 \Return $\textsc{Cand}(A_i)$
 	 	  \caption{\textsc{Column-Selection}}
 	 	  \label{algo_column_retrieval} 
 	 \end{algorithm}

\noindent\textbf{Algorithm}~\ref{algo_column_retrieval} presents the \textsc{column-selection} algorithm. First, it identifies all columns that have a non-empty overlap with the input examples (lines~\ref{crline2}-\ref{crline4}). These columns are then clustered by finding connected components over the hypergraph constructed by the \textsc{discovery engine} (line~\ref{crline5}). To identify the connected components, it uses the discovery \textsc{Neighbors} function. Each cluster is assigned a score that corresponds to the maximum number of examples contained in any column in the cluster (line~\ref{crline7}). This stage supports both \textit{automatic mode} and \textit{interactive mode}. In the \textit{automatic mode}, top-$\theta$ clusters are selected based on their scores (line~\ref{crline8}). 

\mypar{Rationale} Isolating this component in the architecture has two advantages: i) it detects ill-specified queries that lead to too many unnecessary retrieved columns; ii) it offers a point of interaction with \textsc{view-specification} to help users select the right column clusters. Identifying interactive policies to present clusters is outside the scope of this work; we focus on providing the mechanism.

\subsection{Join Graph Search and Materializer}
\label{appendix:c}

The \textsc{Join-Graph-Search} and \textsc{Materializer} components construct candidate PJ-views by joining the tables returned by \textsc{column-selection}. The pseudocode in Algorithm~\ref{algo_view_search} operates in two steps. 



\noindent\underline{1) Join Graph Enumeration (lines~\ref{vsline1}-\ref{vsline10})} enumerates all possible combinations of the columns returned by \textsc{column-selection}, identifies joinable groups of their corresponding tables and find all join graphs in a joinable group. A join graph consists of a group of columns connected via join paths. Given a pair, $c_i, c_j$, of non-joinable columns, no combination of columns involving $c_i$ or $c_j$ can be joined. Non-joinable pairs are cached to skip computation.


\noindent\underline{2) Ranking and Materialization (lines~\ref{vsline11}-\ref{vsline12})} materializes the top-$k$ views as ranked according to the discovery engine score. The discovery engine ranks views according to how well join graphs approximate PK/FK, and according to the size of the join graph; smaller graphs rank higher. Views are materialized incrementally via a materializer built on top of Pandas~\cite{pandas}; it can be adapted to use an embedded database~\cite{sqlite, duckdb} or a processing engine with the \emph{external table} feature~\cite{redshiftexternal} to achieve better performance. 

    

\begin{algorithm}
\scriptsize
\SetKwFunction{AllCombinationOf}{AllCombinationOf} \SetKwInOut{Input}{Input}\SetKwInOut{Output}{Output}
	
	\Input{QBE-style query $\chi$, candidate columns $\textsc{Cand}$,
	expected number of output views $k$
	} 

	\Output{$\mathcal{V}_{PJ}$: list of candidate PJ-views}
	\tcc*[h]{Step 1: Join Graph Enumeration}\\
	$\tau\leftarrow |\textsc{Columns}(\chi)|$\label{vsline1}\\
	$\Gamma\leftarrow \{(c_1,\ldots, c_\tau): \forall c_i\in \textsc{Cand}(A_i)\}$ \\
	$\Gamma_{join}\leftarrow \phi$\\
	 \For{ $r\equiv (c_1,\ldots, c_\tau) \in \Gamma$}
	 {
	 \tcc*[h]{Get join graphs  between source tables of $r$ with $\rho=2$}\\
	 $P\leftarrow
	 \textsc{Generate-Join-Graphs}(r.tables(),2)$ \\
	 \If{$\exists c_i,c_j \in r \mid P(c_i,c_j) = \phi$}{
	    Remove all candidates $r\in \Gamma$ containing columns $c_i$ and $c_j$ such that $P(c_i,c_j)=\phi$  \\
	    Continue\\ 
	 }\Else{
	    $\Gamma_{join}\leftarrow \Gamma_{join} \cup  P$\label{vsline10}
	 }
	 }
	 \tcc*[h]{Step 2: Ranking and Materialization}\\
	 $\Gamma_{join}'\leftarrow $ Select top $k$ join candidates based on join score generated by the discovery engine.\label{vsline11}\\
	 $\mathcal{V}_{PJ}\leftarrow \textsc{materialize-views}(\Gamma_{join}')$\label{vsline12}\\
	
	\Return $\mathcal{V}_{PJ}$
    \caption{\textsc{Join-Graph-Search}}
 	\label{algo_view_search} 
\end{algorithm}

\input{sections/microbenchmarks}

%% file: sections/microbenchmarks.tex
\subsection{Microbenchmarks}
\label{appendix:d}



In this section, we explore the effect of various query and data parameters on \sys{}'s performance.

\subsubsection{\textbf{Varying Discovery Index Quality}}

The quality of the discovery indices built by the Discovery Engine has an effect on the set of candidate PJ-views generated by \sys{} (\textbf{Challenge 2}). We study its effect by modifying the default threshold Aurum uses to compute the hypergraph. In particular, we use thresholds 0.8 (the default), 0.7, 0.6, and 0.5. These thresholds lead to 435, 582, 2143, and 2947 joinable column pairs, respectively. As the thresholds lower, the schema quality worsens, leading to more spurious join paths. These configurations evaluate the effect of schema quality on the discovery of PJ-views over pathless table collections, \ie \textbf{Challenge 2}. As expected, \F~\ref{fig:microbenchmark}(a) shows that the number of join graphs increase as the schema quality worsens. As shown in the previous section, this leads to a higher number of join graphs, and consequently, higher runtimes.


\mypar{Conclusion} When no join path information is available in the schema, inferring the join paths automatically leads to noisy and incorrect ones, which in turn, has an effect on the scalability of the problem (\textbf{Challenge 2}). In conclusion, i) discovering PJ-views on pathless table collections is strictly more difficult than on settings with perfect schemas and; ii) investment in better join path discovery algorithms would highly benefit this problem as well.

\subsubsection{\textbf{Vary the number of Rows in the Query}}
\begin{figure}
  \centering
  \includegraphics[width=\linewidth]{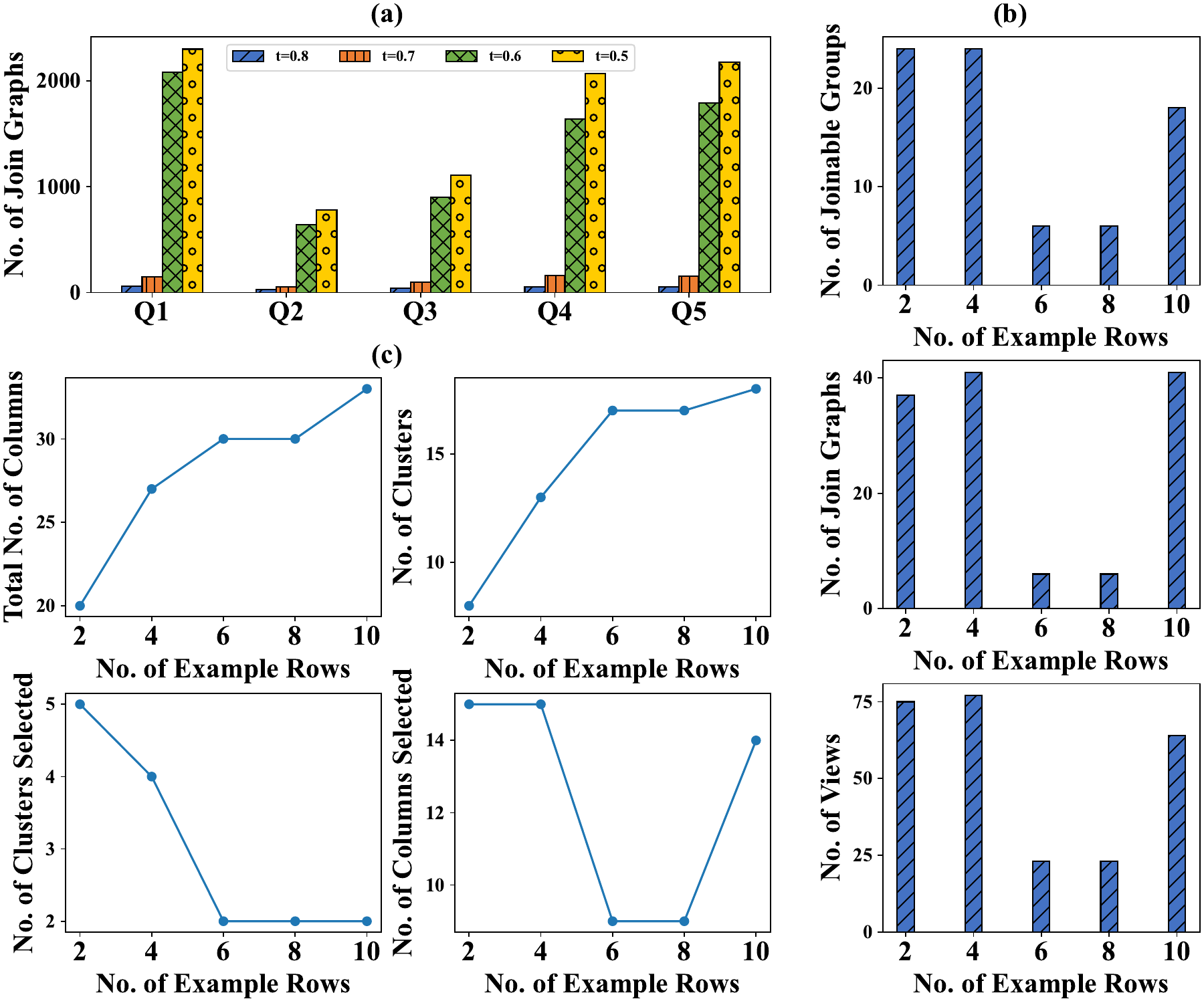}
  \caption{\textbf{(a)} \#join graphs under different $t$ on ChEMBL. Aurum produces 435 (0.8), 582 (0.7), 2143 (0.6), and 2947
  (0.5) joinable column pairs. \textbf{(b)} \#joinable groups, join graphs,
  views of different sample size queries   \textbf{(c)}  \#columns, clusters, selected clusters and
  selected columns of different sample size queries }
  \label{fig:microbenchmark}
\end{figure}

We increase the number of rows inside one query view and observe its effect on the number of joinable groups, join graphs and views (i.e. the search space). 
As shown in \F~\ref{fig:microbenchmark}(b), the relationship between the number of rows and the search space is not monotonous. Increasing the number of rows in a query view can cause the search space to shrink or grow. This is because there are two factors affecting the search space conversely when the number of rows in a query increases. 


\mypar{Enlarge the search space}  \F~\ref{fig:microbenchmark}(c) shows that the total No. of columns before clustering increase as the No. of rows increase and the No. of column clusters will increase as well.

\mypar{Shrink the search space} \F~\ref{fig:microbenchmark}(c) indicates that the number of clusters that \textsc{column-selection} selects will decrease as the No. of examples increases. This is because the score of the ground truth column and its corresponding cluster increases.

\mypar{Conclusion} Intuitively, more rows should lead to fewer candidate PJ-views at the end, and previous work has demonstrated this when schemas are well-formed. But in pathless table collections this is not the case as demonstrated here.

\subsubsection{\textbf{Vary the number of Columns in the query}}
We study the effect of increasing the number of columns in the input query. We choose query with 2, 3 and 4 columns on ChemBL. Unlike varying the number of rows, the results of this experiment are intuitive: higher number of columns in the input lead to higher number of join graphs, candidate PJ-views, and runtime, we do not plot the data for space reasons.